\documentclass[a4paper,12pt]{article}
\usepackage{hyperref}
\usepackage{graphicx,color}
\renewcommand{\theequation}{\arabic{section}.\arabic{equation}}
\begin{document} 
\begin{titlepage}
\null \hfill hep-th/0506112\\
\null \hfill Preprint TU-747  \\
\null \hfill June 2005 \\[4ex]

\begin{center}
{\LARGE On A Superfield Extension of The ADHM Construction 
and ${\cal N}=1$ Super Instantons 
}\\[3em] 
%%%%%%%%%% Here is the title of this Note
%\today

{\large %%%%%%%%%%%%%%%%%% here authors name %%%%%%%%%%%%
Takeo Araki\footnote{
E-mail: araki@tuhep.phys.tohoku.ac.jp
}, \ 
Tatsuhiko Takashima\footnote{
E-mail: takashim@tuhep.phys.tohoku.ac.jp
} \  and \ %\\
Satoshi Watamura\footnote{
E-mail: watamura@tuhep.phys.tohoku.ac.jp
} 
 \\ [2ex]
}
%%%%%%%%%%%%%%%%%%%%%%%%%%%%%% address of the institute %%%%%%%%
Department of Physics \\
Graduate School of Science \\
Tohoku University \\
Aoba-ku, Sendai 980-8578, Japan \\ [2ex]

\vspace{1cm}
%%%%%%%%%%%%%%%%%%%%%%%%%%%%% This is abstruct %%%%%%%%%%%%%%%

\begin{abstract}
 We give a superfield extension of the ADHM
 construction for the Euclidean theory obtained by Wick rotation from 
the Lorentzian four dimensional ${\cal N}=1$ super Yang-Mills theory. 
In particular, we investigate the procedure to guarantee the Wess-Zumino gauge 
for the superfields obtained by the extended ADHM construction, 
and show that the known super instanton configurations are correctly obtained. 
\end{abstract}

\end{center}

\end{titlepage}

%\tableofcontents

%%%%%%%%%%%%%%%%%%%%%%%%%%%%%%%%%%%%%%%%%%%%%%%%%%%%%%%%%%%%%%%%%
\section{Introduction}\setcounter{equation}{0}

Instantons and their effects in field theory 
have been investigated for more than two decades 
(see, for example, \cite{DoHoKhMa,KhMaSl} and references therein) 
and still attract much attention in theoretical physics. 
It is well known that the instanton configurations of the gauge field 
can be obtained by the ADHM construction \cite{AtHiDrMa}. 
In supersymmetric theories, 
there are zero modes of adjoint fermions 
in the instanton background, which naturally introduce 
the superpartner of the bosonic moduli 
called Grassmann collective coordinates (or fermionic moduli). 
The fermion zero modes together with the bosonic configurations 
are called super instantons. 
To give the super instanton solutions, 
superfield extensions of the ADHM construction were proposed in \cite{Sem,Vo}, 
in which the fermionic moduli belong to the same superfield 
containing the corresponding bosonic moduli 
(see also \cite{McAr1}--\cite{DeOg2} for related works). 
Recently it was found that supersymmetric gauge theory defined 
in a kind of deformed superspace, called non(anti)commutative superspace, 
arises in superstring theory 
as a low energy effective theory on D-branes 
%in the presence of constant graviphoton field strength 
\cite{NAC}. 
%In non(anti)commutative space, 
%anticommutators of Grassmann coordinates become non-vanishing. 
%Such non(anti)commutative theory can be realized by deforming 
%the multiplication of superfields. 
%Therefore, 
To obtain instantons in such a deformed theory 
the superfield extension of the ADHM construction would be useful, 
because such a non(anti)commutative theory is realized 
by deforming the multiplication of superfields.

It is recognized, however, that even without deformation 
there exists a gap between 
the superfield extension of the ADHM construction 
and the super instantons obtained by the component formalism. 
We will consider 
the Euclidean theory obtained by Wick rotation from 
the Lorentzian four dimensional ${\cal N}=1$ super Yang-Mills (SYM) theory. 
Hereafter, ``${\cal N}=1$'' stands for the counting of the supercharges 
in the Lorentzian theory before Wick rotation. 
The ADHM construction gives the instanton gauge fields. 
One of the fundamental objects in the ADHM construction is 
the ``zero-dimensional Dirac operator'' 
$ 
\Delta_\alpha (x) 
=
        a_\alpha
        + x_{\alpha\dot{\alpha}} b^{\dot{\alpha}} 
$, where $a$ contains the bosonic moduli parameters. 
Given the zero modes $v$ of $\Delta_\alpha$, 
the instanton gauge fields are constructed 
and written in terms of $a$ and $v$. 
In ${\cal N}=1$ SYM theory, 
there exist fermion zero modes 
coming from the gaugino in the instanton backgrounds. 
It is well known that they are expressed by  
using the ADHM data and 
the fermionic moduli parameters ${\cal M}$. 
When we consider an ${\cal N}=1$ superfield extension of the ADHM construction, 
the operator $\Delta_\alpha (x)$ defined above is extended to 
a chiral superfield 
$\hat{\Delta}_\alpha (y, \theta) = \Delta_\alpha (y) + \theta_\alpha {\cal M}$ 
%including the fermionic moduli ${\cal M}$ 
as was proposed in \cite{Sem,Vo}. 
A superfield obtained from the extended ADHM construction 
as a super instanton configuration 
should be taken in the Wess-Zumino (WZ) gauge 
in order to be compared with the  known results from the component formalism 
and, of course, should be consistent with them 
in its higher components as a superfield. 
So far there is no elaborated procedure to guarantee 
the WZ gauge for the superfields obtained by the extended ADHM construction.

It has been also argued \cite{LuZa} that superfield extensions of the 
ADHM construction for four dimensional Euclidean theories 
exist only for theories 
with extended supersymmetry or complexified gauge group. 
In formulating the extended ADHM construction, 
the authors of refs.\ \cite{Sem,Vo} have introduced 
a counterpart of $\hat\Delta_\alpha(y)$
\footnote{ 
In ref.\ \cite{Vo}, 
it is the transposed matrix of $\hat\Delta_\alpha(y)$. 
}. 
In order to include the purely bosonic ADHM construction, 
this counterpart should coincide with  the complex conjugate of $\Delta_\alpha(x)$ 
when $\theta,\bar{\theta}\rightarrow 0$. 
If we assume that the counterpart is obtained by a kind of conjugation operation, 
such a conjugation maps a chiral (antichiral) superfield in ${\cal N}=1$ theory 
to another chiral (antichiral) superfield. 
In particular, 
this implies that 
the undotted (dotted) spinor $\theta^\alpha$ 
($\bar{\theta}_{\dot{\alpha}}$) is self-conjugate under this conjugation. 
%but we cannot use the usual complex conjugation 
%as the desirable one
%As is well known, 
Since Majorana spinors do not
exist in four dimensional Euclidean space, 
the Grassmann coordinates $\theta^\alpha$ and $\bar{\theta}_{\dot{\alpha}}$ 
in ${\cal N}=1$ theory are necessarily (independent) complex spinors%
\footnote{ 
Similarly $\lambda^\alpha$ and $\bar{\lambda}_{\dot{\alpha}}$ also become 
independent complex spinors. 
Nevertheless, in the path integral, 
Grassmann integration ${\cal D}\lambda$ does not distinguish 
real and complex spinors \cite{NiWa,BeVaNi}, 
so that the fact that $\lambda^\alpha$ and $\bar{\lambda}_{\dot{\alpha}}$ are 
independent complex spinors does not matter as far as we are interested in 
the instanton contribution to the path integral of 
the Lorentzian four dimensional ${\cal N}=1$ theory \cite{DoHoKhMa}. 
} 
and we cannot impose 
the desired reality condition on $\theta$ and $\bar{\theta}$ 
with respect to the complex conjugation. 
%,
%so that in the case we are considering, 
%the complex conjugation should always map the spinors to 
%their complex conjugates (without changing their dottness), rather than themselves.
%; there is no notion of a Majorana spinor 
%in four dimensional Euclidean space ${\bf R}^4$, so that 
%the undotted spinor $\theta^\alpha$ and 
%the dotted spinor $\bar{\theta}_{\dot{\alpha}}$ are independent each other, 
%while reality condition cannot be imposed %in ${\bf R}^4$ 
%unless there are even number of dotted (or undotted) spinors 
%(and it leads to symplectic Majorana spinors). 
%Therefore, in the case we are considering, 
%the complex conjugation should always map the spinors to 
%their complex conjugates (without changing their dottness), rather than themselves. 

In this paper, we would like to 
establish an ${\cal N}=1$ extended ADHM construction, 
i.e.\ a superfield extension of the ADHM construction for 
the Euclidean theory obtained by Wick rotation from 
the Lorentzian four dimensional ${\cal N}=1$ SYM theory. 
We consider ${\rm SU}(n)$ or ${\rm U}(n)$ gauge groups 
for definiteness. 
%Throughout this paper, we concentrate on anti-self dual (ASD) gauge fields. 
As was pointed out in ref.\ \cite{IvLeZu}, for example, 
it is useful to introduce a kind of conjugation (which we denote by ``$\ddagger$'') 
in order to analyze the Euclidean version of the ${\cal N}=1$ theory. 
In terms of the $\ddagger$-conjugation, 
we are allowed to impose ``reality'' conditions for spinors: 
For example, 
$
( \theta_\alpha )^\ddagger 
= 
        \varepsilon^{\alpha\beta} \theta_\beta 
$, 
$
( \bar{\theta}_{\dot{\alpha}} )^\ddagger 
= 
        \varepsilon^{\dot{\alpha}\dot{\beta}} \bar{\theta}_{\dot{\beta}} 
$. 
These conditions imply that ${}^{}{}^\ddagger$ squares to $-1$ 
on spinors and 
this will be one of the characteristic properties of the conjugation. 
With the use of the $\ddagger$-conjugation, 
we investigate the extended ADHM construction, 
where especially care is taken to ensure the WZ gauge for 
the superfields from the extended ADHM construction. 
We show that the ADHM construction can be consistently extended 
in the ${\cal N}=1$ superfield formalism. 
As a result we will see that the obtained gauge potential can be real, 
allowing us to choose ${\rm SU}(n)$ or ${\rm U}(n)$ gauge groups 
rather than their complexification.

This paper is organized as follows. 
In section 2, we review 
the geometric construction of the Euclidean version of 
${\cal N}=1$ SYM theory and 
the ${\cal N}=1$ extended ADHM construction proposed in \cite{Sem,Vo}. 
In section 3, we find the conditions 
in the ${\cal N}=1$ extended ADHM construction 
to give superfields in the WZ gauge 
and show that the resulting superfields are 
actually consistent with known results. 
Section 4 is devoted to conclusions and discussion. 
Our notation and conventions are summarized in appendix A. 
Note that we are working on four dimensional Euclidean space 
but we will use the Lorentzian signature notation of \cite{WeBa}.  
Finally, in appendix B, we describe  
the $\ddagger$-conjugation rules which we have adopted.

%%%%%%%%%%%%%%%%%%%%%%%%%%%%%%%%%%%%%%%%%%%%%%%%%%%%%%%%%%%%%%%%%%%%
\section{The super ADHM construction}\setcounter{equation}{0}
%%%%%%%%%%%%%%%%%%%%%%%%%%%%%%%%%%%%%%%%%%%%%%%%%%%%%%%%%%%%%%%%%%%%
\subsection{The geometric construction of 
${\cal N}=1$ super Yang-Mills\label{s.geo}}

In this subsection, 
we review the geometric construction 
of the Euclidean version of the ${\cal N}=1$ SYM 
theory obtained by Wick rotation 
(see \cite{WeBa,GrSoWe,So} 
and appendix \ref{App:Notation} for our notation and conventions), 
using the superspace formalism. 
In this construction the basic object is the 
connection superfield 
$\phi_A=(\phi_\mu,\phi_\alpha,\phi^{\dot \alpha})$ on
the superspace, 
and the curvature superfield $F_{AB}$ is defined 
(see eq.\ (\ref{eq.fst}) in appendix \ref{App:Notation}). 
$F_{AB}$ satisfies 
the Bianchi identities (\ref{eq:BianchiIdentities}).

The curvature $F_{AB}$ contains many redundant fields 
 and the following constraints are known to be 
the proper ones 
to get rid of the redundant fields 
and to reproduce the multiplet 
of ${\cal N}=1$ SYM theory: 
\begin{equation}
F_{\alpha\beta}=F_{\dot{\alpha}\dot{\beta}}=F_{\alpha\dot{\beta}}=0
. 
\label{eq.constraint}
\end{equation}

We choose the following solution to these constraints: 
\begin{equation} 
\phi_\alpha 
= 
        - e^{-V} D_\alpha e^{V} 
, \quad 
\phi^{\dot{\alpha}} 
= 
        0 
, \quad 
\phi_\mu 
= 
        - \frac i4 \bar{\sigma}_\mu^{\dot{\beta}\beta} 
                \bar{D}_{\dot{\beta}} \phi_{\beta} 
        . 
\label{eq.phi}
\end{equation} 
where $V=V^r T^r$, 
$V^r$ are superfields 
and $T^r$ denote the hermitian generators of ${\rm SU}(n)$ or ${\rm U}(n)$. 
Then the non-trivial curvature superfields 
satisfying the Bianchi identities are written as 
\begin{equation} 
F_{\mu \dot{\alpha}} 
=
      \frac i2 {\cal W}^\beta \sigma_\mu{}_{\beta\dot{\alpha}} 
        , \quad 
F_{\mu \alpha} 
=
      \frac i2 \sigma_\mu{}_{\alpha\dot{\beta}} \bar{\cal W}^{\dot{\beta}} 
        , \quad 
F_{\mu\nu} 
=
    - \frac14 ( \bar{\cal D} \bar{\sigma}_{\mu\nu} \bar{\cal W} 
        - {\cal D} \sigma_{\mu\nu} {\cal W} ) 
, 
\label{eq:Curvatures}
\end{equation} 
where 
\begin{equation} 
{\cal W}^\alpha = W^\alpha 
        , \quad 
\bar{{\cal W}}_{\dot{\alpha}} = e^{-V} \bar{W}_{\dot{\alpha}} e^{V} 
, 
\label{eq:CovChiralSuperfields} 
\end{equation} 
and 
\begin{equation}
W_\alpha 
= 
        - \frac14 \bar{D}_{\dot{\beta}}\bar{D}^{\dot{\beta}} 
                \left(e^{-V} D_\alpha e^V \right)
        , \quad 
\bar{W}_{\dot{\alpha}}
= 
        \frac14 D^\beta D_\beta \left( e^V \bar{D}_{\dot{\alpha}} e^{-V} \right)
. 
\end{equation}

Even after fixing the solution such that $\phi_{\dot{\alpha}}=0$, 
there remains the following gauge freedom: 
\begin{equation}
 e^V \mapsto e^{-i\bar{\Lambda}'} e^{V} e^{i\Lambda} 
, 
\label{eq:ResidualGaugeTransf}
\end{equation}
where $\Lambda$ and $\bar{\Lambda}'$ are an independent 
chiral and anti-chiral superfield in the adjoint representation, respectively. 
This remaining gauge freedom allows us to bring the superfield $V$ 
into the following form (the WZ gauge): 
\begin{equation}
V 
= 
        - \theta\sigma^\mu \bar{\theta} v_\mu (x) 
        + i \theta\theta \bar{\theta} \bar{\lambda} (x) 
        - i \bar{\theta}\bar{\theta} \theta \lambda (x) 
        + \frac12 \theta\theta \bar{\theta}\bar{\theta} D (x) 
. 
\label{eq.fieldV}
\end{equation}
Here all the component fields are complex. 
In this gauge, the remaining gauge freedom of (\ref{eq:ResidualGaugeTransf}) 
is the one with $\Lambda(y,\theta)=\varphi(y)$ and
$\bar{\Lambda}'(\bar y,\bar \theta)=\varphi'(\bar{y})$ 
where $\varphi$ and $\varphi'$ are complex valued functions. 
We require $\varphi'(x) = \varphi(x)$ 
which implies that the transformation (\ref{eq:ResidualGaugeTransf}) 
gives the ordinary gauge transformation laws for the component fields. 
Hereafter, we call $V$ a vector superfield. 
In the WZ gauge, we have 
\begin{eqnarray} 
\phi_\alpha 
&=& 
        \left[ 
        (\sigma^\mu \bar{\theta})_\alpha 
                ( v_\mu + i \theta \sigma_\mu \bar{\lambda} ) 
        - \bar{\theta}\bar{\theta} W_\alpha 
        \right] (y)  
        , \label{eq.phialpha}\\ 
\phi^{\dot{\alpha}} 
&=& 
        0 
        , \\ 
\phi_\mu 
&=& 
        - \frac i2 
        \left[ 
                v_\mu 
                + i \theta \sigma_\mu \bar{\lambda} 
                - \bar{\theta} \bar{\sigma}_\mu W 
        \right] 
        (y) 
\label{eq.phivwz} 
\end{eqnarray}
and 
\begin{eqnarray} 
W_\alpha 
&=& 
        - i \lambda_\alpha (y) 
        + \theta^\gamma \Bigl\{ 
                \varepsilon_{\alpha\gamma} D 
                - i ( \sigma^{\mu\nu} \varepsilon )_{\alpha\gamma} v_{\mu\nu} 
                \Bigr\} (y) 
        + \theta\theta ( \sigma^\mu {\cal D}_\mu \bar{\lambda} )_\alpha (y) 
, 
\label{eq:WinWZ} 
        \\
\bar{W}_{\dot{\alpha}} 
&=& 
        i \bar{\lambda}_{\dot{\alpha}} (\bar{y}) 
        + \bar{\theta}_{\dot{\gamma}} \Bigl\{ 
                \delta^{\dot{\gamma}}_{\dot{\alpha}} D 
                - i ( \bar{\sigma}^{\mu\nu} )^{\dot{\gamma}}{}_{\dot{\alpha}} 
                        v_{\mu\nu} 
                \Bigr\} (\bar{y}) 
        + \bar{\theta}\bar{\theta} 
                ( {\cal D}_\mu \lambda \sigma^\mu )_{\dot{\alpha}} (\bar{y}) 
. 
\end{eqnarray} 
Here $v_{\mu\nu} = \partial_\mu v_\nu - \partial_\nu v_\mu 
+ \frac i2 [ v_\mu , v_\nu ]$ 
and 
${\cal D}_\mu \bar{\lambda} = \partial_\mu \bar{\lambda} 
+ \frac i2 [ v_\mu , \bar{\lambda} ] $.

The Lagrangian is given by 
\begin{equation}
 {\cal L} 
= 
        {-1\over 4 g^2}
        \left(
        \int d^2\theta {\rm tr}W^{\alpha}W_{\alpha}
        + \int d^2\bar{\theta} {\rm tr} \bar{W}_{\dot{\alpha}}
                \bar{W}^{\dot{\alpha}}
        \right) 
= 
        {-1\over g^2} {\rm tr} 
        \left[ 
        - \frac{1}{4} v^{\mu \nu} v_{\mu \nu} 
        - i \bar{\lambda} \bar{\sigma}^{\mu} {\cal D}_{\mu} \lambda 
        + \frac12 D^2 
        \right] 
. 
\end{equation}

%%%%%%%%%%%%%%%%%%%%%%%%%%%%%%%%%%%%%%%%%%%%%%%%%%%%%%%%%%%%%%%%%%%%
%%%%%%%%%%%%%the super ADHM construction%%%%%%%%%%%%%%%%%%%%%%
\subsection{The ${\cal N}=1$ extended ADHM construction\label{sadhm}}
%\footnote{In this section we use the notation of Dorey etal.}

Throughout this paper, we concentrate on anti-self dual (ASD) gauge fields. 
The super ASD instanton configurations satisfy the following equations: 
\begin{equation} 
\star v_{\mu\nu} = - v_{\mu\nu} 
        , \quad 
\sigma^\mu {\cal D}_\mu \bar{\lambda} = 0 
        , \quad 
\lambda_\alpha = 0 
        , \quad 
D = 0 
, 
\label{eq:SuperInstanton} 
\end{equation} 
where 
$\star$ is the Hodge star: 
$\star v^{\mu\nu} = \frac12 \varepsilon^{\mu\nu\rho\sigma} v_{\rho\sigma}$ 
(see appendix \ref{App:Notation} for our convention).

Let us first briefly review the ordinary 
ADHM construction \cite{AtHiDrMa} 
with ${\rm SU}(n)$ (or ${\rm U}(n)$) gauge group. 
The ADHM construction gives the ASD field strength. 
Let us consider a $k$ instanton configuration. 
Define $\Delta_\alpha (x)$ such as 
\begin{equation} 
\Delta_\alpha (x) 
=
        a_\alpha
        + x_{\alpha\dot{\alpha}} b^{\dot{\alpha}} 
\end{equation} 
where $a_\alpha$ and $b^{\dot{\alpha}}$ are constant 
$k \times(n+2k)$ matrices 
and 
$x_{\alpha\dot{\alpha}} 
\equiv 
%       x_\mu \bar{\bf e}^\mu_{\alpha\dot{\alpha}} 
%= 
        i x_{\mu}\sigma^\mu_{\alpha\dot{\alpha}}
$
\footnote{
We include the factor $i$ in the definition of $x_{\alpha\dot{\alpha}}$ 
to ensure $(x_{\alpha\dot{\alpha}})^* = + x^{\dot{\alpha}\alpha}$, 
where ``$*$'' denotes the complex conjugate. 
See appendix \ref{App:Notation}. 
}. 
We assume that $\Delta_\alpha$ has 
maximal rank everywhere except for a finite set of points. 
Its hermitian conjugate 
$\Delta^\dagger{}^\alpha \equiv (\hat{\Delta}_\alpha)^\dagger$ 
is given by 
\begin{equation} 
\Delta^\dagger{}^\alpha (x)
= 
        a^\dagger{}^\alpha 
        + b^\dagger_{\dot{\beta}} x^{\dot{\beta}\alpha} 
. 
\end{equation} 
%Here dagger ${}^{}{}^\dagger$ denotes the Hermitian conjugation 
%\footnote{ 
%In this paper, we will mainly use bar $\bar{}$ 
%to denote dotted spinors and the antichiral coordinates. 
%}. 
%$a_\alpha$ is called the ADHM data. 
%($b^{\dot{\alpha}}$ can be brought into the form 
%$b^{\dot{\alpha}}
%=
%       \left({\bf 0}_{[2k]\times [n]}\ {\bf 1}_{[2k]\times [2k]}\right)$. ) 
Then the gauge field $v_\mu$ is given by 
\begin{equation}
v_\mu
=
        - 2 i v^\dagger \partial_\mu v
        \label{eq.binstv}
, 
\end{equation} 
where $v$ is the set of the normalized zero modes of $\Delta_\alpha$: 
\begin{equation} 
\Delta_\alpha v = 0
        , \quad 
v^\dagger v = {\bf 1}_n 
. 
\end{equation}  
After imposing the bosonic ADHM constraints 
\begin{equation} 
\Delta_\alpha \Delta^\dagger{}^{\beta}\propto\delta_\alpha^\beta
,  
\label{eq:BosADHMConstr}
\end{equation} 
$a_\alpha$ can be identified with the bosonic instanton moduli parameters. 
Finally we find the ASD field strength 
which is implicitly written in terms of the bosonic moduli: 
\begin{equation}
v_{\mu\nu}
=
        8 i v^\dagger b^\dagger \bar{\sigma}_{\mu\nu} f b v 
        \label{eq.binst} 
\end{equation}
where $f$ is defined as the inverse of 
\begin{equation} 
f^{-1} 
= 
        \frac12 \Delta_\alpha \Delta^\dagger{}^\alpha 
. 
\label{eq:finv} 
\end{equation} 

In ${\cal N}=1$ SYM theory, 
there exist zero modes of the adjoint fermion $\bar{\lambda}_{\dot{\alpha}}$ 
in the ASD instanton background (while $\lambda^\alpha=0$) 
and they are written with the use of the ADHM data 
\cite{FermionZeroMode}
as 
\begin{equation}
\bar{\lambda}_{\dot{\alpha}} 
=
         2 i v^\dagger ( b^\dagger_{\dot{\alpha}} f {\cal M} 
                - {\cal M}^\dagger f b_{\dot{\alpha}} ) v 
  \label{eq.finst},
\end{equation}
where ${\cal M}$ is a $k\times (n+2k)$ Grassmann odd matrix 
satisfying the fermionic ADHM constraints 
\begin{equation} 
 {\cal M} \Delta^\dagger_\alpha + \Delta_\alpha {\cal M}^\dagger = 0
. 
\label{eq:FermADHMConstr} 
\end{equation} 
${\cal M}$ can be identified with the fermionic moduli parameters 
after imposing the fermionic ADHM constraints.

The super instanton condition (\ref{eq:SuperInstanton}) can be rewrite 
in the superfield formalism \cite{Sem,Vo} as 
\begin{eqnarray}
&& 
 F_{\mu\dot{\alpha}}=0
\label{eq.SASD} 
        , \\ 
&& 
 \star F_{\mu\nu}=-F_{\mu\nu}
\label{eq.SASD2}
, 
\end{eqnarray} 
where $F$ is the curvature superfield satisfying the Bianchi identity. 
Actually, 
substituting eqs.\ (\ref{eq:SuperInstanton}) 
into $W_\alpha$ in (\ref{eq:WinWZ}),
we find $W_\alpha=0$. This is equivalent to ${\cal W}_\alpha=0$ 
as seen from (\ref{eq:CovChiralSuperfields}). 
Then (\ref{eq.SASD}) and (\ref{eq.SASD2}) 
follow from eq.\ (\ref{eq:Curvatures}). 
The equation (\ref{eq.SASD}) is called the super ASD condition \cite{Vo} 
(note that eq.\ (\ref{eq.SASD}) implies eq.\ (\ref{eq.SASD2})).
%\footnote{
%Note that the super SD condition is also defined by 
%$ F_{\mu\alpha}=0$ (and $\star F_{\mu\nu}=F_{\mu\nu}$).  
%}.
For later use, we give the expression of $F_{\mu\nu}$ 
for the super instanton configuration: 
\begin{equation}
F_{\mu\nu} 
= 
        -\frac{i}{2} 
        \left(
      v^{\mathrm{ASD}}_{\mu\nu}
        - i \theta \sigma^\rho \bar\sigma_{\mu\nu}{\cal D}_\rho \bar\lambda 
        - i\theta\theta \bar{\lambda} \bar{\sigma}_{\mu\nu} \bar{\lambda} 
        \right),
\label{eq.fmninst}
\end{equation}
where $v^{\rm ASD}_{\mu\nu}$ is the ASD part of the field strength $v_{\mu\nu}$. 
Note also that 
the connection superfields (\ref{eq.phialpha})--(\ref{eq.phivwz}) 
for the super instantons take the following forms: 
\begin{equation} 
\phi_\alpha 
= 
        (\sigma^\mu \bar{\theta})_\alpha 
                ( v_\mu + i \theta \sigma_\mu \bar{\lambda} ) 
	(y)  
	, \quad 
\phi_{\dot{\alpha}} 
= 
        0 
	, \quad 
\phi_\mu 
= 
        - \frac i2 
	( 
                v_\mu 
                + i \theta \sigma_\mu \bar{\lambda} 
	) 
        (y) 
\label{eq.phiinst}
. 
\end{equation}

To solve the super ASD condition (\ref{eq.SASD}), 
we introduce a superfield extension of the ADHM construction
along the line of refs.\ \cite{Sem,Vo}. 
By construction, as we will see below, 
the solutions of the extended ADHM construction 
include all the solutions given in the component formalism. 
We define the following quantity which might be interpreted as
a superfield extension of the zero dimensional Dirac operator 
in the WZ gauge:
\begin{equation}
 \hat{\Delta}_\alpha 
                 =\Delta_\alpha (y) + \theta_\alpha {\cal M}
,
\label{eq.SuperDirac}
\end{equation}
where $\Delta_\alpha (y)$ is the zero dimensional Dirac operator in the ordinary
ADHM construction with replacing $x^\mu$ 
by the chiral coordinate $y^\mu = x^\mu + i \theta\sigma^\mu\bar{\theta}$ 
and ${\cal M}$ is a $k\times(n+2k)$ fermionic
matrix which includes the fermionic moduli.
We suppose that $\hat{\Delta}_{\alpha}$ has a maximal rank almost
everywhere as in the ordinary ADHM construction.
%The term almost everywhere means everywhere but a finite set of points.
Also, its counterpart $\widetilde{\hat{\Delta}}{}^\alpha$ is introduced by 
\footnote{ 
In ref.\ \cite{Vo}, 
the counterpart of $\hat{\Delta}_\alpha$ is defined as 
the transposed matrix of $\hat{\Delta}_\alpha$. 
} 
\begin{equation}
\widetilde{\hat{\Delta}}{}^{\alpha}
\equiv 
        \Delta^{\dagger}{}^{\alpha} (y) 
        + \theta^\alpha {\cal M}^\dagger 
        . 
\label{def.conjugation}
\end{equation}
We will show that $\widetilde{\hat{\Delta}}{}^\alpha$ 
can be identified with the $\ddagger$-conjugation of $\hat\Delta_\alpha$ 
in the next section.

%Note that 
The operator $\hat{\Delta}_\alpha$ defined in eq.\ (\ref{eq.SuperDirac}) 
is a chiral superfield without the highest component. 
This can be explained in the following way. 
Since the operator $\Delta_\alpha$ appearing in the ordinary ADHM construction
is related with the Dirac operator reduced to zero dimensional space, 
we expect that its superfield extension $\hat{\Delta}_\alpha$ is related 
to the reduction of the super connection $\phi_\mu$. 
In fact, we see from eq.\ (\ref{eq.phiinst}) that 
the reduction of the connection $\phi_\mu$ for the super instantons 
actually gives a chiral superfield without the highest component 
in the WZ gauge. 
The linearity of $\hat{\Delta}_\alpha$ in terms of $y^\mu$ and $\theta_\alpha$ 
is also suggested by the supersymmetric version of twistor theory
\cite{Sem} 
(see also \cite{Fe,Ma,Vo}).
%The chiral superfield extention correspends to the ``Wess-Zumino gauge'' for the 
%zero-dimensional chiral superfield gauge freedom (\ref{eq.gauge}).

%\item
%Giving the zero dimensional Dirac operator $\hat\Delta_\alpha$,
%we can solve the equation 
%\begin{equation}
%\star F_{\mu\nu}=-F_{\mu\nu}\ ,\label{eq.ASDF}
%\end{equation} 
%in a way parallel to the bosonic case.
%As a result we get the 
%super ADHM constraints as follows:

Another basic object in the ADHM construction is the zero modes 
of $\hat\Delta_\alpha$.
As $\hat\Delta_\alpha$ has $n$ zero modes 
we collect them in a matrix
form:
\begin{equation}
 \hat{\Delta}_\alpha\hat{v}=0
\label{eq.Dv}
\end{equation}
where $\hat{v}$ is an $(n+2k)\times n$ matrix of superfields.
The zero mode superfield
 $\widetilde{\hat{v}}$ of $\widetilde{\hat{\Delta}}{}^\alpha$ 
is also defined by 
\begin{equation} 
\widetilde{\hat{v}} \widetilde{\hat{\Delta}}{}^\alpha = 0
, 
\label{eq:vD} 
\end{equation} 
which is an $n \times (n+2k)$ matrix. 
By construction its lowest component is $v^\dagger$
\footnote{
Note that in ref.\ \cite{Vo} 
the zero mode superfield $\widetilde{\hat v}$ 
of $\widetilde{\hat{\Delta}}{}^\alpha$ is given by the transposed matrix of $\hat{v}$, 
because $\widetilde{\hat{\Delta}}{}^\alpha$ is defined 
as the transposed matrix of $\hat{\Delta}_\alpha$. 
It results in complex field strengths, 
as we will be able to see from $F_{\mu\nu}$ in (\ref{eq.adhmfv}) 
whose lowest component is $- \frac i2 v_{\mu\nu}$. 
In this case, the field strength reads 
$v_{\mu\nu} 
= 
        8 i v^{\rm T} b^{\rm T} \bar{\sigma}_{\mu\nu} f b v 
$ where 
$\hat{f}^{-1} 
\equiv \frac12 \hat{\Delta}_\alpha \hat{\Delta}^{\rm T}{}^\alpha$. 
}. 
We require $\hat{v}$ and $\widetilde{\hat{v}}$ to satisfy 
\begin{equation}
 \widetilde{\hat{v}} \hat{v}=1 
. 
\label{eq.normalization}
\end{equation}

 The matrix $\hat{v}$ defines an embedding of 
the $n$-dimensional vector space into
the $(n+2k)$-dimensional vector space. 
And the projection operator $\hat{v} \widetilde{\hat{v}}$ 
defines the $n$ dimensional vector bundle
embedded in the $n+2k$ dimensional trivial vector bundle.
Using $\hat{v}$ we can pullback 
the trivial
connection of the trivial bundle with rank $(n+2k)$,
and get the non-trivial connection of 
the vector bundle with rank $n$ as follows:
%construct a connection $\phi$ from $\hat{v}$
\begin{equation}
 \phi =-\widetilde{\hat{v}} d\hat v.\label{eq.connectionphi}
\end{equation}
where $d$ is exterior derivative of superspace (see appendix \ref{App:Notation}).
The connection $\phi$ defines the curvature
\begin{equation}
 F 
= 
        d\phi+\phi\phi 
%= 
%       d\left(d\hat{v}^\sharp\hat{v}\hat{v}^\sharp\right)\hat{v} 
%= 
%       d\hat{v}^\sharp(1-\hat{v}\hat{v}^\sharp)d\hat{v} 
= 
        \widetilde{\hat{v}} 
                d \widetilde{\hat{\Delta}}{}^{\alpha}\hat K_\alpha{}^{\beta}
        d\hat{\Delta}_\beta\hat{v}
        ,
\label{eq:ADHM:Curvature2-form}
\end{equation}
where
\begin{equation}
\hat K^{-1}{}_\alpha{}^{\beta}
\equiv 
 \hat\Delta_\alpha \widetilde{\hat\Delta}{}^{\beta} 
= 
        \Delta_\alpha \Delta^\dagger{}^\beta 
        + \theta^\gamma 
                \Bigl( \Delta_\alpha \delta_\gamma^\beta {\cal M}^\dagger 
                + \varepsilon_{\alpha\gamma} {\cal M} \Delta^\dagger{}^\beta \Bigr) 
        + \theta\theta \Bigl( \frac12 \delta_\alpha^\beta 
                {\cal M} {\cal M}^\dagger \Bigr) 
\end{equation}
and $\hat{K}_\alpha{}^\beta$ is defined such that 
$
\hat{K}^{-1}{}_\alpha{}^{\beta} \hat{K}{}_\beta{}^\gamma 
= 
        \hat{K}_\alpha{}^{\beta} \hat{K}^{-1}{}_\beta{}^\gamma  
= 
        \delta_{\alpha}^\gamma {\bf 1}_k 
$. 
To obtain the expression (\ref{eq:ADHM:Curvature2-form}), 
we have used the completeness condition:
\begin{equation}
\hat v \widetilde{\hat{v}} = {\bf 1}_{n+2k} -\widetilde{\hat{\Delta}}{}^{\alpha}
 \hat K_\alpha{}^{\beta}\hat\Delta_\beta.
\end{equation}
The coefficient of the $2$-form $F$ defines the superfield $F_{AB}$:
\begin{equation}
 F_{AB}=-\widetilde{\hat{v}} D_{[A} \widetilde{\hat\Delta}{}^{\alpha}
        \hat K_\alpha{}^{\beta} D_{B\}}\hat{\Delta}_\beta\hat{v}
. 
        \label{eq.2formcoefficients}
\end{equation}

If $\hat K$ commutes with
the Pauli matrix, the field strength $F_{\mu\nu}$ becomes 
ASD as in the bosonic ADHM construction although it is a superfield now. 
This is 
equivalent to the condition 
$\hat\Delta_\alpha \widetilde{\hat\Delta}{}^{\beta}\propto\delta_{\alpha}^{\beta}$ 
and thus 
\begin{equation}
\hat K^{-1}{}_\alpha{}^{\beta} %= \hat\Delta_\alpha \widetilde{\hat\Delta}{}^{\beta}
=\delta_\alpha^{\beta}\hat{f}^{-1}\label{eq.diagonal}
\end{equation}
where 
\begin{equation} 
\hat f^{-1} 
\equiv 
        \frac12 \hat{\Delta}_\alpha \widetilde{\hat\Delta}{}^\alpha 
\label{eq:fhatinv}
\end{equation} 
is a $k\times k$ matrix superfield.
There exists $\hat f$ because we have assumed that 
$\hat\Delta_\alpha$ has maximal rank. 
The above condition (\ref{eq.diagonal}) leads to 
both the bosonic ADHM constraints (\ref{eq:BosADHMConstr}) 
and the fermionic ones (\ref{eq:FermADHMConstr}).  
When eq.\ (\ref{eq.diagonal}) holds, i.e., 
the parameters in $\hat\Delta_\alpha$ are satisfying
both bosonic and fermionic ADHM constraints,
we obtain from eq.(\ref{eq.2formcoefficients}) 
the ASD curvature superfield (\ref{eq.SASD2}) 
and another (non-trivial) one 
in terms of the ADHM quantities: 
\begin{eqnarray}
 F_{\mu\nu}
&=& 
        4 \widetilde{\hat{v}} b^\dagger \bar{\sigma}_{\mu\nu} \hat{f} b \hat{v} 
\label{eq.adhmfv}
        , \\ 
F_{\mu\alpha} 
&=& 
        \frac i2 \sigma_\mu{}_{\alpha\dot{\beta}} 
        \left\{ 
        - 2 \widetilde{\hat{v}} (
                b^\dagger{}^{\dot{\beta}} \hat{f} {\cal M} 
                - {\cal M}^\dagger \hat{f} b^{\dot{\beta}} 
                ) \hat{v} 
   -8 \bar{\theta}_{\dot{\gamma}} 
                \widetilde{\hat{v}} ( 
                b^\dagger{}^{\dot{\beta}} \hat{f} b^{\dot{\gamma}} 
                + b^\dagger{}^{\dot{\gamma}} \hat{f} b^{\dot{\beta}} 
                ) \hat{v}
        \right\} 
. 
\end{eqnarray} 
We can check that the curvature superfields 
satisfy the covariant constraints (\ref{eq.constraint}) 
and the super ASD condition (\ref{eq.SASD}). 
$F_{\dot\alpha\dot\beta}
=F_{\alpha\dot\beta}=0$ and $F_{\mu\dot\alpha}=0$ are simply a consequence of 
the definition of $\hat\Delta_\alpha$ as 
a chiral extension of $\Delta_\alpha$. 
Note that 
$F_{\alpha\beta}=0$ is checked with the use of 
the constraint (\ref{eq.diagonal}) and 
\begin{equation} 
D_\beta \hat{\Delta}_\alpha 
= 
        \varepsilon_{\alpha\beta} 
                ( {\cal M} 
        + 4\bar{\theta}_{\dot{\beta}} b^{\dot{\beta}} ) 
        , \quad 
D_\beta \widetilde{\hat{\Delta}}{}^\alpha 
=
        \delta^\alpha_\beta 
                ( {\cal M}^\dagger 
                + 4 b^\dagger_{\dot{\beta}} 
                        \bar{\theta}^{\dot{\beta}} ) 
                . 
\label{eq:DDelta} 
\end{equation} 
and the fact that $F_{\alpha\beta}$ 
is symmetric with respect to $\alpha$ and $\beta$. 
This implies that all the constraints and the super ASD condition are satisfied by
the connection (\ref{eq.connectionphi}), 
with $\hat v$ being the zero modes of %the zero dimensional Dirac operator 
$\hat\Delta_\alpha$ obeying the constraint (\ref{eq.diagonal}).

So far we have reviewed an ${\cal N}=1$ generalization of the
ADHM construction to the superfield formalism and 
have shown that the extended ADHM construction 
formally gives the ASD connection $\phi$ in the superfield formalism.
The component fields of super instantons in the WZ gauge 
can be found without knowing the higher 
components of $\hat{v}$ and $\hat{f}$. 
They can be obtained by using the following relations: 
\begin{equation} 
v_{\mu\nu} = 2 i F_{\mu\nu} | 
        , \quad 
\lambda_\alpha = i W_\alpha | 
        , \quad 
\bar{\lambda}_{\dot{\alpha}} = - i \bar{{\cal W}}_{\dot{\alpha}} | 
        , \quad 
D = - \frac12 D^\alpha W_\alpha | 
, 
\end{equation} 
where $|$ indicates that we take $\theta=\bar\theta=0$ 
(see also \cite{McAr2}).
We find that $\bar{\cal W}$ is given by 
\begin{equation}
\bar{\cal W}^{\dot{\alpha}} 
=
   - 2 \widetilde{\hat{v}} (
                b^\dagger{}^{\dot{\alpha}} \hat{f} {\cal M} 
                - {\cal M}^\dagger \hat{f} b^{\dot{\alpha}} 
                ) \hat{v}
   - 8 \bar{\theta}_{\dot{\gamma}} 
                \widetilde{\hat{v}} ( 
                b^\dagger{}^{\dot{\alpha}} \hat{f} b^{\dot{\gamma}} 
                + b^\dagger{}^{\dot{\gamma}} \hat{f} b^{\dot{\alpha}} 
                ) \hat{v}
\label{eq.adhmfw}
\end{equation}
since 
$\bar{\cal W}^{\dot{\alpha}} 
=
   \frac i2 \bar{\sigma}^\mu{}^{\dot{\alpha}\beta} F_{\mu\beta}
$, 
and $W^\alpha=0$ from ${\cal W}^\alpha 
=
 \frac i2 \bar{\sigma}^\mu{}^{\dot{\beta}\alpha} F_{\mu\dot{\beta}}$. 
$F_{\mu\nu}$ is given in (\ref{eq.adhmfv}). 
Thus, we obtain 
\begin{eqnarray} 
&& 
v_{\mu\nu}
=
        8 i v^\dagger b^\dagger \bar{\sigma}_{\mu\nu} f b v 
        , 
\label{eq:ASD:FieldStrength} 
        \\ 
&& 
\bar{\lambda}_{\dot{\alpha}} 
=
         2 i v^\dagger ( b^\dagger_{\dot{\alpha}} f {\cal M} 
                - {\cal M}^\dagger f b_{\dot{\alpha}} ) v 
        , 
\label{eq:ASD:LambdaBar} 
        \\ 
&& 
\lambda_\alpha=D=0 
, 
\label{eq:ASD:LambdaD}
\end{eqnarray} 
where $f$ is the lowest component of $\hat{f}$ 
and given by the inverse of (\ref{eq:finv}) 
for the case under consideration. 
The above fields precisely coincide with the super instanton configuration 
 in eqs. (\ref{eq.binst}) and (\ref{eq.finst}). 
Note that here we have only used the lowest components of $\hat{v}$ and $\hat{f}$, 
hence we do not need to solve the equation (\ref{eq.Dv}). 
However, 
we will solve eq.\ (\ref{eq.Dv}) in the next section 
and determine the full form of the connection $\phi_\mu$,
in order to confirm that this formalism can 
lead to the correct superfield extension of the ordinary ADHM construction. 
For this end, the information of the higher components of
$\hat v$ and $\hat{f}$ is necessary 
and this will be a non-trivial check of the consistency of our extension. 
%To find the higher components of $\hat{v}$ is also important especially 
%when we consider the deformation of the superspace, 
%where the anti-commutator of $\theta$ do not
%vanish and higher components of superfields emerge in the lower
%components of the product superfields. 

%%%%%%%%%%%%%%%%%%%%%%%%%%%%%%%%%%%%%%%%%%%%%%%%%%%%%%%%%%%%%%%%%%%%%%%%%%%%
\section{
Consistency of the extended ADHM construction}\setcounter{equation}{0}

\subsection{The conjugation} 

As we discussed in the previous section, 
to formulate the extended ADHM construction 
a kind of conjugation is needed, 
which maps the superfields $\hat{\Delta}_\alpha$ and $\hat{v}$ 
to $\widetilde{\hat{\Delta}}{}^\alpha$ and $\widetilde{\hat{v}}$,
 respectively. 
In order to determine the full form of the connection superfield $\phi_\mu$, 
it is found that we need to consider 
such a kind of conjugation also for other quantities. 
This ``conjugation'' operation 
should map a chiral superfield in ${\cal N}=1$ theory 
to another chiral superfield. 
Thus the $\ddagger$-conjugation satisfies at least the following properties: 
It maps the undotted (dotted) spinor $\theta^\alpha$ 
($\bar{\theta}_{\dot{\alpha}}$) to itself and 
the chiral coordinate $y^\mu$ is invariant under this conjugation. 
In addition, the ``conjugation'' of the quantities 
appearing in the purely bosonic ADHM construction 
should reduce to the ordinary complex (hermitian) conjugation.

A conjugation suitable for our purpose 
(which we denote as ``$\ddagger$'') 
has been considered in the literature (for example, \cite{IvLeZu}). 
The $\ddagger$-conjugation can be characterized by the following properties: 
\begin{eqnarray} 
&& 
A^\ddagger{}^\ddagger = (-)^{|A|} A
        , 
\label{eq:ddagger1} 
        \\ 
&& 
(y^\mu)^\ddagger = y^\mu 
        , 
\label{eq:ddagger2}
        \\ 
&& 
B^\ddagger = B^\dagger 
\label{eq:ddagger3} 
, 
\end{eqnarray} 
where $A$ denotes any quantity appearing in this paper 
($|A|$ denotes its Grassmann oddness) 
and $B$ appearing in the purely bosonic ADHM construction. 
In addition, we require that 
$\theta_\alpha$ and $\bar{\theta}_{\dot{\alpha}}$ satisfy 
the following ``reality'' conditions: 
\begin{equation} 
( \theta_\alpha )^\ddagger 
= 
        \theta^\alpha 
        , \quad 
( \bar{\theta}_{\dot{\alpha}} )^\ddagger 
= 
        \bar{\theta}^{\dot{\alpha}} 
. 
\label{eq:Reality:Theta} 
\end{equation} 
We also introduce for each quantity 
its counterpart in the ``anti-holomorphic'' sectors 
with respect to the $\ddagger$-conjugation, 
which we indicate by a tilde ``$\, \widetilde{\ ^{}}\, $''. 
The rules we have adopted are given in appendix \ref{App:Conjugation}%
\footnote{In particular, the $\ddagger$-conjugation of a product 
of two fermionic quantities $A,B$ is defined by
$
(A B)^\ddagger 
= 
        - B^\ddagger A^\ddagger
$. }.

We require the ``reality'' of the supercharges 
under the $\ddagger$-conjugation as well as the Grassmann coordinates: 
\begin{equation} 
( Q_\alpha )^\ddagger 
= 
        Q^\alpha 
        , \quad 
( \bar{Q}_{\dot{\alpha}} )^\ddagger 
= 
        \bar{Q}^{\dot{\alpha}} 
, 
\label{eq:Reality:Q} 
\end{equation} 
Then we can check that the supersymmetry algebra is self-conjugate 
under the $\ddagger$-conjugation. 

Moreover, with the use of the $\ddagger$-conjugation 
(and the above ``reality'' conditions), 
we can obtain the real gauge potential $v_\mu$  
in the Euclidean version of the ${\cal N}=1$ SYM theory. 
We can impose the ``reality'' condition on the superfield $V$ 
in (\ref{eq.fieldV}) as a superfield equation: 
\begin{equation} 
V^\ddagger 
= 
        - V
        . 
\label{eq:Reality:V} 
\end{equation} 
This condition implies 
\footnote{ 
The auxiliary field $D$ is hermitian in the Lorentzian theory. 
On the other hand, according to the rule in appendix \ref{App:Conjugation}, 
$D^\ddagger = - D$ implies that $D$ is anti-hermitian. 
Since we are considering the Euclidean theory, 
the hermiticity of $D$ is not necessary. 
}
\begin{equation} 
v_\mu^\dagger = v_\mu 
        , \quad 
( \lambda_\alpha )^\ddagger 
= 
        \lambda^\alpha 
        , \quad 
( \bar{\lambda}_{\dot{\alpha}} )^\ddagger 
= 
        \bar{\lambda}^{\dot{\alpha}} 
        , \quad 
D^\ddagger 
= 
        - D
        . 
\label{eq:Reality:Components} 
\end{equation} 
Note that in our convention we have 
$(\theta\theta)^\ddagger = \theta \theta$, 
$(\bar{\theta}\bar{\theta})^\ddagger = \bar{\theta}\bar{\theta}$ 
and 
$(\theta \sigma^\mu \bar{\theta})^\ddagger = - \theta \sigma^\mu \bar{\theta}$. 
In the WZ gauge, 
as we have described in section \ref{s.geo}, 
the gauge transformation (\ref{eq:ResidualGaugeTransf}) reduces to 
\begin{equation}
 e^V \mapsto e^{-i\varphi(\bar{y})} e^{V} e^{i\varphi(y)}  
, 
\end{equation}
where $\varphi$ is a complex valued function. 
After the imposition of (\ref{eq:Reality:V}), 
we find that $\varphi(y)^\ddagger = \varphi(y)$ is required. 
This leads to $\varphi(x)^\dagger = \varphi(x)$ 
and the condition $v_\mu^\dagger = v_\mu$ is preserved 
even after the gauge transformation. 
Thus we conclude that the gauge potential $v_\mu$ is real 
and the gauge group is ${\rm SU}(n)$ or ${\rm U}(n)$ 
rather than their complexification. 

Also, the connection $\phi$ and the curvature $F$ have 
the following gauge freedom  
\begin{equation} 
\phi \mapsto X^{-1} \phi X - X^{-1} d X 
        , \quad 
F \mapsto X^{-1} F X 
, 
\label{eq:XGaugeTransf}
\end{equation} 
where $X$ is a generic superfield extension of the ordinary gauge transformation. 
In our choice of the solution (\ref{eq.phi}) to the covariant constraints, 
$(\phi_\mu)^\ddagger = - \phi_\mu$ holds, 
which follows from eq.\ (\ref{eq:Reality:Components}) 
and can be checked with the use of 
the expression (\ref{eq.phivwz}) in the WZ gauge. 
Therefore, the superfield $X$ should satisfies 
\begin{equation} 
X^\ddagger X = X X^\ddagger = {\bf 1}_n 
\label{eq:XX=1} 
\end{equation} 
or $X^\ddagger = X^{-1}$ 
after imposition of (\ref{eq:Reality:V}). 

Now let us apply the $\ddagger$-conjugation 
to the extended ADHM construction introduced in the previous section. 
According to the $\ddagger$-conjugation rules, we can check that 
$(\Delta_\alpha)^\ddagger$ 
coincides with $\widetilde{\hat{\Delta}}{}^\alpha$ 
defined in (\ref{def.conjugation}): 
\begin{equation} 
(\hat\Delta_\alpha)^\ddagger
= 
        \widetilde{\hat\Delta}{}^\alpha 
. 
\end{equation} 
Also, 
$\widetilde{\hat v}$ appearing in (\ref{eq:vD}) should also be identified 
with the ``anti-holomorphic counterpart'' of $\hat v$ 
with respect to the $\ddagger$-conjugation; 
from eqs.\ (\ref{eq.connectionphi}) and (\ref{eq:ADHM:Curvature2-form})
we find that in the extended ADHM construction 
the $X$ transformation (\ref{eq:XGaugeTransf}) is realized when 
$\hat v$ and its counterpart $\widetilde{\hat v}$ appearing in (\ref{eq:vD}) 
transform as 
\begin{equation} 
\hat{v} \mapsto \hat{v}' 
= 
        \hat{v} X
        , \quad 
\widetilde{\hat{v}} \mapsto \widetilde{\hat{v}'}
= 
        X^\ddagger \widetilde{\hat{v}} 
, 
\label{eq:XGaugeTransf:v} 
\end{equation} 
where $X^\ddagger = X^{-1}$. 
This leads us to require $\widetilde{\hat v}$ 
to be the $\ddagger$-conjugate of $\hat v$: 
\begin{equation} 
\hat v^\ddagger 
= 
        \widetilde{\hat v} 
. 
\end{equation} 
According to these identifications, 
the $\ddagger$-conjugation rules tells us that 
the curvature superfield $F_{\mu\nu}$ in (\ref{eq.adhmfv}) 
given by the extended ADHM construction 
enjoys the following property: 
\begin{equation} 
(F_{\mu\nu})^\ddagger 
= 
        - F_{\mu\nu} 
. 
\end{equation} 
Here we have used $\hat{f}^\ddagger = \hat{f}$ which follows 
from (\ref{eq:fhatinv}). 
Since the lowest component of $F_{\mu\nu}$ is $-\frac i2 v_{\mu\nu}$, 
the above equation tells us that 
\begin{equation} 
v_{\mu\nu}^\dagger 
= 
        v_{\mu\nu} 
, 
\end{equation} 
which further implies $v_\mu^\dagger = v_\mu$. 
This means that the gauge potential $v_{\mu}$ 
given by the extended ADHM construction can be kept and shown to be real 
if we adopt the $\ddagger$-conjugation.

\subsection{The Wess-Zumino gauge and the component fields} 

In this subsection, 
we show that the extended ADHM construction and the $\ddagger$-conjugation 
can give the correct super instanton configurations. 
As we have discussed in the previous subsection, 
$\widetilde{\hat\Delta}{}^\alpha$ and $\widetilde{\hat v}$ 
in section \ref{sadhm} are identified with the ``anti-holomorphic'' counterpart 
of $\hat\Delta_\alpha$ and $\hat v$ with respect to the $\ddagger$-conjugation, 
respectively. 
Keeping this identification in mind, 
we can use all the equations in section \ref{sadhm} without any changes. 
We derive the 
normalized zero modes $\hat v$ of $\hat\Delta_\alpha$ 
by solving eq.\ (\ref{eq.Dv}) 
and determine the connection $\phi_\mu$.

The zero mode $\hat{v}$ of $\hat\Delta_\alpha$ could be a general superfield. 
Given such a zero mode $\hat{v}$, 
we can find another gauge equivalent zero mode $\hat{v}'$ by 
the transformation (\ref{eq:XGaugeTransf:v}). 
We can always expand $\hat{v}$ in terms of $\bar{\theta}$ as 
\begin{equation} 
\hat{v} (y, \theta, \bar{\theta})
= 
        \hat{u} (y, \theta) 
        + \bar{\theta}_{\dot{\alpha}} \hat{u}'^{\dot{\alpha}} (y, \theta) 
        + \bar{\theta}\bar{\theta} \hat{u}'' (y, \theta) 
        , 
\end{equation} 
where $\hat{v} (y, \theta, \bar{\theta})$ denotes 
the expression of the zero modes in the chiral basis. 
$\hat{u}$ should be normalized as $\widetilde{\hat{u}} \hat{u} = {\bf 1}_n$ 
because of (\ref{eq.normalization}).  
Because $\hat{\Delta}$ is a chiral superfield, 
$\hat{\Delta}_\alpha \hat{v}$ is expanded 
in terms of $\bar{\theta}$ as 
\begin{equation} 
\hat{\Delta}_\alpha \hat{v} (y, \theta, \bar{\theta})
= 
        \hat{\Delta}_\alpha \hat{u} (y, \theta) 
        + \bar{\theta}_{\dot{\alpha}} 
                \hat{\Delta}_\alpha \hat{u}'^{\dot{\alpha}} (y, \theta) 
        + \bar{\theta}\bar{\theta} \hat{\Delta}_\alpha \hat{u}'' (y, \theta) 
\end{equation} 
and the zero mode equation leads to the following equations: 
\begin{equation} 
\hat{\Delta}_\alpha \hat{u} (y, \theta) = 0 
        , \quad 
\hat{\Delta}_\alpha \hat{u}'^{\dot{\alpha}} (y, \theta) = 0 
        , \quad 
\hat{\Delta}_\alpha \hat{u}'' (y, \theta) = 0 
. 
\end{equation} 
Once the 
normalized
 chiral superfield $\hat{u}$ 
is found, 
the fields $\hat{u}'^{\dot{\alpha}}$ and $\hat{u}''^{\dot{\alpha}}$ 
can be obtained with the use of $\hat{u}$ by 
\begin{equation} 
\hat{u}'^{\dot{\alpha}} 
= 
        \hat{u} A^{\dot{\alpha}} 
        , \quad 
\hat{u}'' 
= 
        \hat{u} B 
        , 
\end{equation} 
where $A^{\dot{1}}$, $A^{\dot{2}}$ and $B$ are 
arbitrary $n\times n$ matrices of chiral superfields. 
Therefore the zero mode $\hat{v}$ can always be written as 
\begin{equation} 
\hat{v} (y, \theta, \bar{\theta})
= 
        \hat{u} (y, \theta) {\bf A} (y, \theta, \bar{\theta}) 
\end{equation} 
where 
\begin{equation} 
{\bf A} (y, \theta, \bar{\theta}) 
\equiv 
        {\bf 1}_n (y, \theta) 
        + \bar{\theta}_{\dot{\alpha}} A^{\dot{\alpha}} (y, \theta) 
        + \bar{\theta}\bar{\theta} B (y, \theta) 
, 
\end{equation} 
and ${\bf A}^\ddagger = {\bf A}^{-1}$ from the normalization conditions 
of $\hat{v}$ and $\hat{u}$. 
The superfield ${\bf A}$ is eliminated 
by the $X$ gauge transformation (\ref{eq:XGaugeTransf:v}) 
with $X={\bf A}^{-1}$. 
Therefore, we can always restrict the zero mode superfield $\hat{v}$ 
to a chiral superfield.

When $\hat{v}$ and $\widetilde{\hat{v}}$ are chiral superfields, 
we find from (\ref{eq.connectionphi}) that 
$
\phi_{\dot{\alpha}}
= 
        0 
, 
$
and also that $\phi_\mu$ is a chiral superfield 
($
\bar{D}_{\dot{\alpha}} \phi_\mu = 0 
$) 
because eq.\ (\ref{eq.connectionphi}) tells us that $\phi_\mu$ is given by 
\begin{equation} 
\phi_\mu (y, \theta) 
= 
        - \widetilde{\hat{v}}  
                {\partial \over \partial y^\mu} \hat{v} (y, \theta) 
. 
\end{equation} 
These facts are consistent with the connection superfields for the super instantons
(\ref{eq.phiinst}). 
In the following, we show that we can actually find $\hat v$ 
which is a chiral superfield 
and correctly gives the connection superfields (\ref{eq.phiinst}). 

Considering $\phi_{\alpha}$, 
we find non-trivial necessary conditions to determine such $\hat v$. 
In the ADHM construction, $\phi_\alpha$ can be written in the chiral basis as  
\begin{equation} 
\phi_\alpha 
= 
        - \widetilde{\hat{v}} D_\alpha \hat{v} 
= 
        - \widetilde{\hat{v}} 
                {\partial \over \partial \theta^\alpha} \hat{v} (y, \theta) 
        + 2 i (\sigma^\mu \bar{\theta})_{\alpha} \phi_\mu (y, \theta)
. 
\end{equation} 
On the other hand, from (\ref{eq.phiinst}) 
we see that $\phi_\alpha$ for the super instanton solutions should satisfy 
\begin{equation} 
\phi_\alpha 
= 
        2 i (\sigma^\mu \bar{\theta})_{\alpha} \phi_\mu (y, \theta)
        , 
\label{eq:phialpha_phimu}
\end{equation} 
in the WZ gauge. 
As a result, it should hold that 
\begin{equation} 
\widetilde{\hat{v}} 
                {\partial \over \partial \theta^\alpha} \hat{v} (y, \theta) 
= 
        0 
        , 
\label{eq:v:NecessaryCond} 
\end{equation} 
which is a necessary condition for $\hat{v}$ to be in the WZ gauge. 
In the component language, 
denoting the zero mode chiral superfield $\hat{v}$ as 
\begin{equation} 
\hat{v} (y, \theta) 
\equiv 
        v^{(0)} (y) 
        + \theta^\gamma v^{(1)}_\gamma (y) 
        + \theta\theta v^{(2)} (y) 
        , 
\end{equation} 
the above condition gives 
\begin{eqnarray} 
&& 
\widetilde{v}^{(0)} v^{(1)}_\alpha 
= 
        0 
        \label{eq:v:NecessaryCond1} 
        , \\
&&
\widetilde{v}^{(0)} v^{(2)}\delta^\beta{}_\alpha
= 
        \frac12 \widetilde{v}^{(1)}{}^\beta v^{(1)}_\alpha
        \label{eq:v:NecessaryCond2} 
        , \\ 
&& 
\widetilde{v}^{(1)}_\alpha v^{(2)} 
= 
         \widetilde{v}^{(2)} v^{(1)}_\alpha  
        \label{eq:v:NecessaryCond3} 
. 
\end{eqnarray} 
In addition to this, the $\theta\theta$-component of 
the ASD connection $\phi_\mu$ should vanish in the WZ gauge: 
\begin{equation} 
\phi_\mu |_{\theta\theta} 
= 
        \widetilde{v}{}^{(2)} \partial_\mu v^{(0)} 
        + \widetilde{v}{}^{(0)} \partial_\mu v^{(2)} 
        - \frac12 \widetilde{v}{}^{(1)}{}^\gamma \partial_\mu v^{(1)}_\gamma 
= 0 
. 
\end{equation} 
We can use these conditions to determine $\hat{v}$ in the WZ gauge.

The zero dimensional Dirac equation (\ref{eq.Dv}) can be written as 
\begin{equation} 
        \Delta_\alpha v^{(0)} 
        + \theta^\gamma 
                \Bigl( 
                \Delta_\alpha v^{(1)}_\gamma 
                + \varepsilon_{\alpha\gamma} {\cal M} v^{(0)} 
                \Bigr) 
        + \theta\theta 
                \Bigl( 
                \Delta_\alpha v^{(2)} 
                + \frac12 {\cal M} v^{(1)}_\alpha 
                \Bigr) 
= 
        0 
. 
\end{equation} 
With a given $v^{(0)}$ that satisfies 
$\Delta_\alpha v^{(0)} = 0$ and $\widetilde{v}^{(0)} v^{(0)} ={\bf 1}_n$, 
this equation is solved by 
\begin{eqnarray} 
v^{(1)}{}^\gamma 
&=& 
        \widetilde{\Delta}{}^\gamma f {\cal M} v^{(0)} 
        + v^{(0)} \chi^\gamma 
\label{eq:v1withchi}
        , \\ 
v^{(2)} 
&=& 
        - \frac12 \widetilde{\Delta}{}^\gamma f {\cal M} 
        \Bigl( 
        \widetilde{\Delta}{}_\gamma f {\cal M} v^{(0)} 
        + v^{(0)} \chi_\gamma 
        \Bigr) 
        + v^{(0)} s 
\label{eq:v2withs}
, 
\end{eqnarray}
where $\chi^\alpha$ and $s$ are an arbitrary $n\times n$ 
fermionic and bosonic matrix, respectively. 
With the use of the $\ddagger$-conjugation rules, 
we also find 
\begin{equation} 
\widetilde{\hat{v}} (y, \theta) 
= 
        \widetilde{v}^{(0)} (y) 
        + \theta^\gamma \widetilde{v}^{(1)}_\gamma (y) 
        + \theta\theta \widetilde{v}^{(2)} (y) 
        , 
\end{equation} 
where $\widetilde{v}{}^{(0)} $ satisfies 
$\widetilde{v}{}^{(0)} \tilde{\Delta}{}^\alpha = 0$ 
and $\widetilde{v}^{(0)} v^{(0)} = {\bf 1}_n$, and  
\begin{eqnarray} 
\widetilde{v}{}^{(1)}_\gamma  
&=& 
        - \widetilde{v}{}^{(0)} \widetilde{{\cal M}} f \Delta_\gamma 
        + \widetilde{\chi}{}_\gamma \widetilde{v}{}^{(0)} 
, 
\\ 
\widetilde{v}{}^{(2)} 
&=& 
        \frac12 
        \Bigl( 
        - \widetilde{v}{}^{(0)} 
                \widetilde{{\cal M}} f \Delta^\gamma 
        + \widetilde{\chi}{}^\gamma \widetilde{v}{}^{(0)} 
        \Bigr)
                \widetilde{{\cal M}} f \Delta_\gamma 
        + \widetilde{s} \widetilde{v}{}^{(0)} 
. 
\end{eqnarray} 
To obtain this expression, we have used 
$( \widetilde{\Delta}^\gamma )^\ddagger = \Delta_\gamma$, 
$f^\ddagger = f$%
. 
We can check that $\widetilde{\hat{v}}$ given by the above components 
actually satisfies 
$\widetilde{\hat{v}} \widetilde{\hat{\Delta}}{}^\alpha = 0$%
. 

Next we will determine $\hat{v}$ in the WZ gauge. 
Substituting (\ref{eq:v1withchi}) and (\ref{eq:v2withs}) 
into the necessary conditions 
(\ref{eq:v:NecessaryCond1}) and (\ref{eq:v:NecessaryCond2}), 
we find 
\begin{equation} 
\chi_\alpha = 0
        , \quad 
s 
= 
        \frac12 \widetilde{v}^{(0)} \widetilde{{\cal M}} f {\cal M} v^{(0)} 
        .  
\end{equation}  
The third condition (\ref{eq:v:NecessaryCond3}) 
is fulfilled by these $\chi_\alpha$ and $s$. 
The resulting zero mode $\hat{v}$ in the WZ gauge is 
\begin{equation} 
\hat{v} 
= 
        v^{(0)} 
        + \theta^\gamma \Bigl( \widetilde{\Delta}_\gamma f {\cal M} v^{(0)} \Bigr) 
        + \theta\theta 
                \Bigl( \frac12 \widetilde{{\cal M}} f {\cal M} v^{(0)} \Bigr) 
        , 
\label{eq:vinWZ}
\end{equation} 
and its conjugate is 
\begin{equation} 
\widetilde{\hat{v}} 
= 
        \widetilde{v}^{(0)} 
        + \theta^\gamma \Bigl( - \widetilde{v}^{(0)} \widetilde{{\cal M}} 
                f \Delta_\gamma \Bigr) 
        +\theta\theta 
                \Bigl( \frac12 \widetilde{v}^{(0)} \widetilde{{\cal M}} f {\cal M} \Bigr) 
        . 
\end{equation} 
Here we have used the fermionic ADHM constraint 
${\cal M} \widetilde{\Delta}_\gamma = - \Delta_\gamma \widetilde{{\cal M}}$. 
We can check that this $\hat{v}$ satisfies the normalization condition.

We can explain the relation between 
the zero mode $\hat{v}$ given by (\ref{eq:v1withchi}) and (\ref{eq:v2withs}), 
and the one in the WZ gauge (\ref{eq:vinWZ}), 
in terms of the $X$ gauge transformation. 
Note that 
even if the zero mode superfield $\hat v$ is taken to be a chiral superfield, 
there is a remaining $X$ gauge freedom (\ref{eq:XGaugeTransf:v}) 
by a chiral superfield $X(y,\theta)$. 
Denoting the chiral superfield $X$ as 
\begin{equation} 
X (y, \theta) 
= 
        X^{(0)} (y) 
        + \theta^\gamma X^{(1)}_\gamma (y) 
        + \theta\theta X^{(2)} (y) 
, 
\end{equation} 
we find that the components of $\hat{v}$ transform as 
\begin{eqnarray} 
&& 
v^{(0)} 
\mapsto 
        v^{(0)} X^{(0)} 
        , \\
&& 
v^{(1)}_\alpha 
\mapsto 
        v^{(1)}_\alpha X^{(0)} + v^{(0)} X^{(1)}_\alpha 
        , \\
&& 
v^{(2)} 
\mapsto 
        v^{(2)} X^{(0)} + v^{(0)} X^{(2)} 
        - \frac12 v^{(1)}{}^\gamma X^{(1)}_\gamma 
. 
\end{eqnarray} 
It is easy to check that the gauge transformation of the former $\hat{v}$ 
by the following $X$ 
%(which satisfies the condition (\ref{eq:XX=1})) 
gives $\hat{v}$ in the WZ gauge (\ref{eq:vinWZ}): 
\begin{equation} 
X 
= 
        {\bf 1}_n  
        + \theta^\gamma \Bigl( - \chi_\gamma \Bigr) 
        + \theta\theta 
                \Bigl(
                - s 
                - \frac12 \chi^\gamma \chi_\gamma 
                + \frac12 \widetilde{v}^{(0)} \widetilde{{\cal M}} f {\cal M} v^{(0)} 
                \Bigr) 
. 
\end{equation}

Fixing the gauge to the one in which $\hat{v}$ is given as in (\ref{eq:vinWZ}), 
the $X$ symmetry breaks down 
to the ordinary ${\rm SU}(n)$ (or ${\rm U}(n)$) gauge symmetry; 
now only $X (y, \theta) = X^{(0)} (y)$ is allowed, 
thus $X^\ddagger = X^{-1}$ means 
$X^{(0)}(y){}^\ddagger = X^{(0)}(y){}^{-1}$,  
which further implies that $X^{(0)}(x){}^\dagger = X^{(0)}(x){}^{-1}$.

We check that the above $\hat{v}$ correctly gives 
the connection $\phi$ in the WZ gauge of our solution (\ref{eq.phi}). 
For the instanton solutions in the WZ gauge, 
$\phi_\mu$ should be 
$
\phi_\mu 
= 
        - \frac i2 
        ( 
                v_\mu 
                + i \theta \sigma_\mu \bar{\lambda} 
        ) 
        (y) 
$
with $v_\mu$ and $\bar{\lambda}$ being (\ref{eq.binstv}) 
and (\ref{eq.finst}) respectively. 
In fact, substituting the above $\hat{v}$ and $\widetilde{\hat{v}}$ 
into eq.\ (\ref{eq.connectionphi}), we find that the connection now becomes 
\begin{equation} 
\phi_\mu 
= 
        - \widetilde{v}{}^{(0)} \partial_\mu v^{(0)} 
        + i \theta^\gamma \sigma_\mu{}_{\gamma\dot{\beta}} 
                \widetilde{v}^{(0)} ( - \widetilde{b}^{\dot{\beta}} f {\cal M} 
                - \widetilde{{\cal M}} f b^{\dot{\beta}} ) v^{(0)} 
. 
\end{equation} 
Here the $\theta\theta$-component vanishes due to the relation
\begin{equation} 
\partial_\mu f 
= 
        f \Delta^\gamma \partial_\mu \widetilde{\Delta}_\gamma f 
\label{eq:Delf} 
\end{equation} 
which can be proved with the use of the bosonic ADHM constraints and 
the canonical forms of $b$ and $\widetilde{b}$. 
Writing 
$v^{(0)}=v$, we see that $\widetilde{v}^{(0)}=v^\dagger$, 
$\widetilde{{\cal M}}={\cal M}^\dagger$ and 
$\widetilde{b}= - b^\dagger$ follow
from the definition of the $\ddagger$-conjugation 
(see appendix \ref{App:Conjugation}), and 
 the above $\phi_\mu$ coincides with the connection in the WZ
gauge:
\begin{equation} 
\phi_\mu 
= 
        - \frac i2 \left[ 
        - 2 i v^\dagger \partial_\mu v 
        + i \theta^\gamma \sigma_\mu{}_{\gamma\dot{\beta}} 
                \left\{ 2 i v^\dagger ( b^{\dagger \dot{\beta}} f {\cal M} 
                - {\cal M}^\dagger f b^{\dot{\beta}} ) v \right\} \right]
. 
\end{equation} 
Thus $\phi_\mu$ correctly reproduces the gauge field 
(\ref{eq.binstv}) and the fermion zero mode (\ref{eq.finst}) 
of super instantons. 
Note that 
$\phi_\alpha$ satisfies (\ref{eq:phialpha_phimu}) 
and $\phi_{\dot{\alpha}} =0$, by construction.

With the use of (\ref{eq:vinWZ}), 
we can also check that  
$F_{\mu\nu}$ in (\ref{eq.adhmfv}) gives the correct component fields 
of (\ref{eq.fmninst}). 
To verify this, we need $\hat f$: 
>From the eqs.(\ref{eq.SuperDirac}) and (\ref{eq.diagonal}), 
$\hat f^{-1}$ is given by 
\begin{equation} 
\hat{f}^{-1} 
= 
        f^{-1}  
        + \theta^\gamma \Delta_\gamma \widetilde{{\cal M}}
        + \frac12 \theta\theta {\cal M} \widetilde{{\cal M}}  
\end{equation}
and its inverse $\hat f$ is found to be 
\begin{equation}
\hat{f} 
= 
        f \Bigl[ 
        f^{-1} 
        - \theta^\gamma \Delta_\gamma \widetilde{\cal M} 
        - \frac12 \theta\theta {\cal M} 
                ( 1 - \widetilde{\Delta}^\gamma f \Delta_\gamma ) \widetilde{\cal M} 
        \Bigr] f 
        . 
\label{eq.fhat}
\end{equation}

Finally, let us 
discuss the dimension of the moduli space
in the super field formalism of the instanton solution.
The curvature $F$ is invariant under the following 
${\rm GL}(k) \times {\rm U}(n+2k)$ 
global symmetry transformation 
\begin{equation} 
\hat{\Delta}_\alpha 
\rightarrow 
        G \hat{\Delta}_\alpha \Lambda 
        , \quad 
\hat{f} 
\rightarrow 
        G \hat{f} G^\dagger 
        , \quad 
\hat{v} 
\rightarrow 
        \Lambda^{-1} \hat{v} 
, 
\end{equation} 
where $G \in {\rm GL}(k)$ and $\Lambda \in {\rm U}(n+2k)$. 
After fixing $b$ in the canonical form, 
the global symmetry breaks down to ${\rm U}(n) \times {\rm U}(k)$ 
as in the purely bosonic ADHM construction. 
Therefore, the number of the bosonic parameters contained 
in $\hat \Delta_\alpha$ is 
$4nk$ as the
usual bosonic case after imposing the 
bosonic ADHM constraints (\ref{eq:BosADHMConstr}). 
The number of fermionic parameters
 contained in $\cal M$ is $2k(n+2k)$. 
There is no additional symmetry and thus 
the number of fermionic parameters is reduced 
simply by the fermionic ADHM constraints (\ref{eq:FermADHMConstr}) 
and is $2kn$.

\section{Conclusions and discussion}\setcounter{equation}{0} 

In this paper, we have studied the ${\cal N}=1$ superfield extension of 
the ADHM construction. 
We found a procedure to guarantee the WZ gauge for 
the superfields from the extended ADHM construction. 
Using the $\ddagger$-conjugation 
which enables us to impose a kind of reality condition, 
we have shown that the extended ADHM construction correctly gives 
super instanton configurations 
for the theory with ${\rm SU}(n)$ or ${\rm U}(n)$ gauge group, 
rather than their complexification. 

It would be interesting to apply the extended ADHM construction 
to field theories in non(anti)commutative superspace \cite{NAC}. 
The deformed super instanton configurations in such theories \cite{DInst}
can be obtained by replacing the product of superfields 
appearing in the extended ADHM construction by the Moyal type $*$-product. 
In order to find the component fields of super instantons, 
to guarantee the WZ gauge is especially important,  
and this step can be accomplished by using the $\ddagger$-conjugation. 
A detailed analysis will appear in a forthcoming paper \cite{ATW2}.

{\bf Acknowledgments}:
The authors would like to thank U.~Carow-Watamura for discussions.
T.~A. is very grateful to the members of 
the Particle Theory and Cosmology Group 
at Tohoku University for their support and hospitality, 
and would also like to thank K.~Ohashi for useful comments.
This research is partly supported by Grant-in-Aid for Scientific
Research from the Ministry of Education, Culture,
Sports, Science and Technology, Japan No.\ 13640256 and 13135202. 
T.~A. is supported by 
Grant-in-Aid for Scientific Research in Priority Areas (No.\ 14046201) 
from the Ministry of Education, Culture, Sports, Science 
and Technology.

\renewcommand{\theequation}{\Alph{section}.\arabic{equation}}
%%%%%%%%%%%%%%%%%%%%%%%%%%%%%%%%%%%%%%%%%%%%%%%
\appendix
%%%%%%%%%%%notation%%%%%%%%%%%%%%%%%%%%%%%%%%%%%%%%%%%%%%%%%%%%%%%%
%%%%%%%%%%%%%%%%%%%%%%%%%%%%%%%%%%%%%%%%%%%%%%%%%%%%%%%%%%%%%%%%%%%
\section{Notation and conventions}\setcounter{equation}{0} 
\label{App:Notation} 

By Wick rotation $x_{\rm E}^0=x_{\rm E}{}_0=i x^0$, 
we obtain the Euclidean theory with the metric 
$\eta^{\mu\nu}=\eta_{\mu\nu}={\rm diag}(1,1,1,1)$ 
where $\mu,\nu=0,1,2,3$. 
We drop the subscript ``${}_{\rm E}$''. 

The antisymmetric $\varepsilon$-tensors for the spinor indices are defined by 
\begin{equation}
\varepsilon_{21}=\varepsilon^{12} 
= \varepsilon_{\dot{2}\dot{1}} = \varepsilon^{\dot{1}\dot{2}} = + 1. 
\end{equation}

We use the following sigma matrices: 
\begin{equation}
\sigma^\mu = \sigma_\mu
\equiv 
        (-i{\bf 1}, \sigma^i)
        , \quad 
\bar{\sigma}^\mu = \bar{\sigma}_\mu
\equiv 
        (-i{\bf 1}, -\sigma^i)
, 
\end{equation}
where $\sigma^i$ are the Pauli matrices and 
$
\bar{\sigma}^{\mu}{}^{\dot{\alpha}\alpha}
= 
        \varepsilon^{\dot{\alpha}\dot{\beta}}
        \varepsilon^{\alpha\beta}\sigma_{\beta\dot{\beta}}^\mu
$ 
holds. 
The Lorentz generators are 
\begin{equation}
\sigma^{\mu\nu} 
\equiv  
        \frac14 (\sigma^\mu \bar{\sigma}^\nu - \sigma^\nu \bar{\sigma}^\mu) 
, \quad 
\bar{\sigma}_{\mu\nu} 
\equiv \frac14 ( \bar{\sigma}_\mu \sigma_\nu - \bar{\sigma}_\nu \sigma_\mu )
, 
\end{equation} 
where 
\begin{equation} 
\sigma^{\mu\nu} 
= 
        \frac {1}{2} \varepsilon^{\mu\nu\lambda\rho} 
                \sigma_{\lambda\rho} 
, \quad 
\bar{\sigma}^{\mu\nu} 
= 
        - \frac {1}{2} \varepsilon^{\mu\nu\lambda\rho} 
                \bar{\sigma}_{\lambda\rho} 
, \quad 
\varepsilon^{0123} = \varepsilon_{0123} \equiv -1  
. 
\end{equation} 
We define 
\footnote{
Note that here the factor $i$ is contained so that these definitions 
to be the same as those using the quaternion basis in the spinor notation. 
However we do not use a bar ``$\ \bar{}\ $'' 
to indicate the conjugation of a quaternion, 
since $\bar y_\mu$
is used to denote the antichiral coordinates.
} 
\begin{equation} 
x_{\alpha\dot{\beta}}
\equiv 
%       x_\mu \bar{\bf e}^\mu_{\alpha\dot{\beta}} 
%= 
        i x_\mu \sigma^\mu_{\alpha\dot{\beta}} 
        , \quad 
x^{\dot{\beta}\alpha}  
\equiv 
%       x_\mu {\bf e}_\mu^{\dot{\beta}\alpha} 
%= 
        i x_\mu \bar{\sigma}_\mu^{\dot{\beta}\alpha} 
. 
\end{equation}

The supersymmetry algebra is 
\begin{eqnarray} 
&&
\{ Q_\alpha, \bar{Q}_{\dot{\beta}} \} = 2 \sigma^\mu_{\alpha\dot{\beta}} P_\mu 
, 
\\ 
&& 
\{ Q_\alpha, Q_\beta \} 
= 
        \{  \bar{Q}_{\dot{\alpha}}, \bar{Q}_{\dot{\beta}} \}
        = 0 
        , \quad 
        [ P_\mu , \cdot \ ] = 0
        . 
\end{eqnarray} 
In the coordinate representation, the generators are written as 
\begin{equation}
Q_{\alpha} 
=
        {\partial\over\partial \theta^\alpha}
        - i \sigma_{\alpha\dot{\alpha}}^\mu \bar{\theta}^{\dot{\alpha}} 
                {\partial\over\partial x^\mu}
                , \quad 
\bar{Q}_{\dot{\alpha}}
= 
        - {\partial\over\partial \bar{\theta}^{\dot{\alpha}}}
        + i \theta^{\alpha} \sigma_{\alpha\dot{\alpha}}^\mu 
                {\partial\over\partial x^\mu}
        , \quad 
P_\mu = i {\partial\over\partial x^\mu} 
. 
\end{equation}
The supercovariant derivatives are 
\begin{equation}
D_{\alpha} 
= 
        {\partial\over\partial \theta^\alpha}
        + i \sigma_{\alpha\dot{\alpha}}^\mu \bar{\theta}^{\dot{\alpha}} 
                {\partial\over\partial x^\mu}
                , \quad 
\bar{D}_{\dot{\alpha}}
= 
        -{\partial\over\partial \bar{\theta}^{\dot{\alpha}}}
        - i \theta^{\alpha} \sigma_{\alpha\dot{\alpha}}^\mu 
                {\partial\over\partial x^\mu}
. 
\end{equation}

The chiral and antichiral coordinates are 
\begin{equation}
 y^{\mu}=x^{\mu}
+i\theta^{\alpha}\sigma^{\mu}_{\alpha\dot{\alpha}}
\bar{\theta}^{\dot{\alpha}}
, \quad 
 \bar{y}^{\mu}=x^{\mu}
-i\theta^{\alpha}\sigma^{\mu}_{\alpha\dot{\alpha}}
\bar{\theta}^{\dot{\alpha}}
. 
\end{equation}
We also define 
\begin{equation} 
y_{\alpha\dot{\alpha}}
\equiv 
%       y_\mu \bar{\bf e}^\mu_{\alpha\dot{\alpha}} 
%= 
        i y_\mu \sigma^\mu_{\alpha\dot{\alpha}} 
        , \quad 
y^{\dot{\alpha}\alpha}  
\equiv 
%       y_\mu {\bf e}_\mu^{\dot{\alpha}\alpha} 
%= 
        i y_\mu \bar{\sigma}_\mu^{\dot{\alpha}\alpha} 
. 
\end{equation} 

Define the following basis in the cotangent space of the superspace: 
\begin{equation} 
(e^A) 
= 
        (e^\mu, e^\alpha, e_{\dot{\alpha}}) 
        , \qquad 
e^\mu 
= 
        d y^\mu - 2 i d \theta \sigma^\mu \bar{\theta} 
        , 
\quad  
e^\alpha 
= 
        d \theta^\alpha 
        , 
\quad 
e_{\dot{\alpha}} 
= 
        d \bar{\theta}_{\dot{\alpha}} 
. 
\end{equation}  
Then the exterior derivative $d$ is expanded 
in the $\{ e^A \}_{A=\mu, \alpha, \dot{\alpha}}$-basis as 
\begin{equation}
d = e^A D_A 
, 
\end{equation}
where 
\begin{equation}
(D_A)
= 
        \left({\partial / \partial y^\mu}, D_\alpha, D^{\dot{\alpha}} \right)  
. 
\end{equation} 

The connection is a Lie algebra valued one-form: 
\begin{equation} 
\phi 
= 
        e^A \phi_A 
        , \qquad 
\phi_A
= 
        \phi_A^r i T^r
,
\end{equation} 
($T^r$ are the hermitian generators in a certain representation of the gauge group). 
We will define 
\begin{equation} 
\phi_\mu |_{\theta=\bar{\theta}=0} 
\equiv  
        - \frac i2 v_\mu
, 
\end{equation} 
where $v_\mu = v_\mu^r T^r$ is the gauge potential field.

The curvature two-form $F$ is given by 
\begin{equation} 
F 
= 
        d \phi + \phi \wedge \phi 
= 
        \frac12 e^A \wedge e^B F_{BA} 
. 
\end{equation} 
We find 
\begin{equation} 
F_{AB} 
= 
        D_A \phi_B -(-)^{ab} D_B \phi_A 
        - [ \phi_A , \phi_B \} 
        + T_{AB}{}^C \phi_C 
        , 
\label{eq.fst}
\end{equation} 
where $a$ and $b$ denote the Grassmann oddness of $\phi_A$ and $D_B$,
 respectively, 
$[ \cdot , \cdot \}$ is a supercommutator 
and $T_{AB}{}^C$ is the torsion defined by 
$ 
d e^C 
= 
        \frac12 e^A e^B T_{BA}{}^C 
$ 
whose non-vanishing elements are 
$
T_{\alpha\dot{\beta}}{}^\mu = T_{\dot{\beta}\alpha}{}^\mu 
= 2i \sigma^\mu_{\alpha\dot{\beta}}
$. 
We can see 
\begin{equation} 
F_{\mu\nu} |_{\theta=\bar{\theta}=0} 
= 
        - \frac i2 v_{\mu\nu} 
, 
\end{equation} 
where $v_{\mu\nu} = v_{\mu\nu}^r T^r $ is the usual field strength, 
$
v_{\mu\nu} 
= 
        \partial_\mu v_\nu - \partial_\nu v_\mu + \frac i2 [ v_\mu, v_\nu ] 
$. 

The curvature two-form $F$ satisfies 
the Bianchi identity: 
$
{\cal D} F 
= 
        d F - \phi \wedge F + F \wedge \phi 
= 
        0 
        , 
$
or 
\begin{equation} 
\frac12 e^A e^B e^C 
        \left( 
        D_C F_{BA} - [ \phi_C , F_{BA} \} 
        + \frac12 T_{CB}{}^D F_{DA} 
        + \frac12 T_{CA}{}^D F_{DB} 
        \right) 
= 
        0 
        . 
        \label{eq:BianchiIdentities} 
\end{equation} 

The ``zero dimensional Dirac operator'' in the extended ADHM construction 
is defined by 
\begin{equation} 
\hat{\Delta}_\alpha
        {}_{[k]\times[n+2k]} 
= 
        \Delta_\alpha 
        + \theta_\alpha {\cal M} 
        , 
\end{equation} 
where 
\begin{equation} 
\Delta_\alpha 
\equiv 
        a_\alpha
        + y_{\alpha\dot{\beta}} b^{\dot{\beta}}
\end{equation} 
and 
\begin{eqnarray}
&& 
a_{\alpha}
        {}_{[k]\times[n+2k]}  
\equiv 
        \Bigl(\matrix{
                \omega_{\alpha}
                        {}_{[k]\times[n]}
                & 
                (a'{}_{\alpha\dot{\beta}})
                        {}_{[k]\times[2k]}
        }\Bigr) 
= 
        \Bigl(\matrix{
                (\omega_{\alpha}{}^i{}_u) 
                & 
                (a'{}_{\alpha\dot{1}}{}^i{}_j) 
                & 
                (a'{}_{\alpha\dot{2}}{}^i{}_j)
        }\Bigr) 
        , \nonumber\\
&& 
{\cal M} 
        {}_{[k]\times[n+2k]}  
\equiv 
        \Bigl(\matrix{
                \mu
                        {}_{[k]\times[n]}
                & 
                ({\cal M}'{}_{\dot{\beta}}) 
                        {}_{[k]\times[2k]}
        }\Bigr) 
= 
        \Bigl(\matrix{
                (\mu^i{}_u) 
                & 
                ({\cal M}'{}_{\dot{1}}{}^i{}_j) 
                & 
                ({\cal M}'{}_{\dot{2}}{}^i{}_j) 
        }\Bigr) 
, 
\end{eqnarray} 
with $u=1,\dots ,n$ and $i,j=1,\dots , k$. 
Note that we write 
$a'{}_{\alpha\dot{\beta}}{}^j{}_i 
= 
        i a'^\mu{}^j{}_i \sigma_\mu{}_{\alpha\dot{\beta}}
$. 
The canonical form of $b$ is defined as 
\begin{equation}
(b^{\dot{\alpha}})
= 
        \left(\matrix{
        b^{\dot{1}} \cr b^{\dot{2}}
        }\right)
= 
        \left(\matrix{
        {\bf 0}_{[k]\times [n]} & {\bf 1}_k & {\bf 0}_k \cr 
        {\bf 0}_{[k]\times [n]} & {\bf 0}_k & {\bf 1}_k 
        }\right) 
. 
\end{equation}

%%%%%%%%%%%%%extended conjugation operation%%%%%%%%%%%%%%%%%%%%%%%%
%%%%%%%%%%%%%%%%%%%%%%%%%%%%%%%%%%%%%%%%%%%%%%%%%%%%%%%%%%%%%%%%%%%
\section{The $\ddagger$-conjugation rules}\setcounter{equation}{0}
\label{App:Conjugation} 

In this appendix, we will describe a map 
which we call the $\ddagger$-conjugation. 
There are many possibilities to choose a set of rules 
satisfying eqs.\ (\ref{eq:ddagger1}), 
(\ref{eq:ddagger2}) and (\ref{eq:ddagger3}) 
for imposing ``reality'' conditions like eq.\ (\ref{eq:Reality:Theta}). 
The rules described here are one of the possible ways 
which turns out to be sufficient for our purpose. 

Let $A_{ij}$, $A^\alpha{}_{ij}$ and $\bar{A}^{\dot{\alpha}}{}_{ij}$ 
are matrices in the ``holomorphic sector'' 
in terms of the $\ddagger$-conjugation. 
We require the existence of the matrices in the ``anti-holomorphic sector'' 
$\widetilde{A}_{ij}$, $\widetilde{A}{}^\alpha{}_{ij}$ 
and $\widetilde{\bar{A}}{}^{\dot{\alpha}}{}_{ij}$, respectively. 
The matrices in the holomorphic sector and those  
in the anti-holomorphic sector  
are related by the $\ddagger$-conjugation as follows%
: 
For a (bosonic or fermionic) matrix $A$, 
\begin{equation} 
(A)^\ddagger 
= 
        \widetilde{A} 
        , \quad 
(\widetilde{A})^\ddagger 
= 
        (-)^{|A|} A     
        .
\end{equation} 
For matrices with one spinor index, 
\begin{eqnarray} 
&& 
(A_\alpha)^\ddagger 
= 
        \widetilde{A}{}^\alpha
        , \quad 
(\widetilde{A}{}^\alpha)^\ddagger 
= 
        (-)^{|A|} A_\alpha
\label{eq:ddagger:undotted} 
        \\ 
&& 
(\bar{A}_{\dot{\alpha}})^\ddagger 
=  
        \widetilde{\bar{A}}{}^{\dot{\alpha}}
        , \quad 
(\widetilde{\bar{A}}{}^{\dot{\alpha}})^\ddagger 
= 
        (-)^{|\bar{A}|} \bar{A}_{\dot{\alpha}}
\label{eq:ddagger:dotted} 
. 
\end{eqnarray} 
As a consequence, we have 
$A^{\ddagger\ddagger} 
= 
        (-)^{|A|} A
$, 
$(A^\alpha)^{\ddagger\ddagger}  
        = 
                (-)^{|A|} A^\alpha$ 
and 
$(\bar{A}^{\dot{\alpha}})^{\ddagger\ddagger}  
        = 
                (-)^{|\bar{A}|} \bar{A}^{\dot{\alpha}}
$. 
The $\ddagger$-conjugation of a product 
of two (bosonic or fermionic) quantities is defined by 
\begin{equation} 
(A B)^\ddagger 
= 
        (-)^{|A||B|} B^\ddagger A^\ddagger
        . 
\end{equation}
As long as the quantities in the anti-holomorphic sector 
are not specified, it is always possible to consider 
such a map for any quantity. 

For the quantities appearing in the (non-supersymmetric) pure 
gauge theory and also in the ordinary ADHM construction, 
we require that the $\ddagger$-conjugation coincides 
with the usual hermitian (complex) conjugation $\dagger$: 
For example, 
\begin{eqnarray} 
&& 
v_\mu{}^\ddagger 
= 
        v_\mu{}^\dagger 
        , \quad 
(\omega_\alpha)^\ddagger 
= 
        (\omega_\alpha)^\dagger 
        = \omega^{\dagger}{}^{\alpha} 
        , \quad 
(b^{\dot{\alpha}})^\ddagger 
= 
        (b^{\dot{\alpha}})^\dagger 
        = b^\dagger{}_{\dot{\alpha}} 
        , \\ 
&& 
(\varepsilon_{\alpha\beta})^{\ddagger} 
= 
        (\varepsilon_{\alpha\beta})^* 
= 
        - \varepsilon^{\alpha\beta} 
        , \quad 
(\sigma^\mu_{\alpha\dot{\beta}})^\ddagger 
= 
        (\sigma^\mu_{\alpha\dot{\beta}})^* 
= 
        - \bar{\sigma}^\mu{}^{\dot{\beta}\alpha} 
, \quad \mbox{etc.}
\end{eqnarray} 
 The changes of the upper and the lower spinor indices 
under the $\ddagger$-conjugation 
in (\ref{eq:ddagger:undotted}) and (\ref{eq:ddagger:dotted}) 
have been arranged to allow us this identification. 
Then the corresponding quantities in the anti-holomorphic sector 
are specified in accordance with the hermitian (complex) conjugation rule 
for each quantity in the holomorphic sector. 
In particular, we should notice that 
$\widetilde{b}_{\dot{\alpha}} = - b^\dagger_{\dot{\alpha}}$ 
in our convention. 

For fermionic quantities with one spinor index, 
we are allowed to impose a ``reality'' condition 
in terms of the $\ddagger$-conjugation. 
This is accomplished by identifying the anti-holomorphic quantity 
with such a spinor itself. 
For example, we may require 
such spinors $\psi^\alpha$ and $\bar{\psi}^{\dot{\alpha}}$ 
to satisfy $\widetilde{\psi}^\alpha = \psi^\alpha$ 
and $\widetilde{\bar{\psi}}{}^{\dot{\alpha}}=\bar{\psi}^{\dot{\alpha}}$, 
that is, 
\begin{equation} 
(\psi_\alpha)^\ddagger = \psi^\alpha
        , \quad 
(\bar{\psi}_{\dot{\alpha}})^\ddagger 
        = \bar{\psi}^{\dot{\alpha}} 
        . 
\end{equation} 
The ``reality'' conditions (\ref{eq:Reality:Theta}) 
on $\theta$ and $\bar{\theta}$ are defined such that  
$y_\mu^\ddagger = y_\mu$ holds. 

For the other quantities appearing in this paper, 
we identify the anti-holomorphic quantity 
with the hermitian (complex) conjugate, 
if it exists, of each holomorphic quantity like 
\begin{equation} 
\widetilde{{\cal M}} = {\cal M}^\dagger 
, 
\end{equation} 
or with the quantity deduced from the known rules, if we can compute: 
For example, the anti-holomorphic counterpart of $\hat{\Delta}_\alpha$ 
can be computed by using the rules for $\Delta_\alpha$, $\theta_\alpha$ 
and ${\cal M}$, such that 
\begin{equation} 
\widetilde{\hat{\Delta}}{}^\alpha 
= 
        \Delta^\dagger{}^\alpha 
        + \theta^\alpha {\cal M}^\dagger 
        . 
\end{equation} 
We should note that the hermitian conjugation 
and the $\ddagger$-conjugation for fermionic quantities 
do not commute in general. 
One of such examples is ${\cal M}$: 
We have 
$({\cal M}^\ddagger)^\dagger = {\cal M}^\dagger{}^\dagger = {\cal M}$, 
while 
$({\cal M}^\dagger)^\ddagger 
= \widetilde{{\cal M}}^\ddagger = (-)^{|{\cal M}|} {\cal M}$.


\begin{thebibliography}{99}

\bibitem{DoHoKhMa}
N.~Dorey, T.~J.~Hollowood, V.~V.~Khoze and M.~P.~Mattis,
``The calculus of many instantons,''
Phys.\ Rept.\  {\bf 371} (2002) 231
[arXiv:hep-th/0206063]; 
%%CITATION = HEP-TH 0206063;%%

\bibitem{KhMaSl} 
V.~V.~Khoze, M.~P.~Mattis and M.~J.~Slater,
``The instanton hunter's guide to supersymmetric ${\rm SU}(N)$ gauge theory,''
Nucl.\ Phys.\ B {\bf 536} (1998) 69
[arXiv:hep-th/9804009].
%%CITATION = HEP-TH 9804009;%%

\bibitem{AtHiDrMa}
M.~F.~Atiyah, N.~J.~Hitchin, V.~G.~Drinfeld and Y.~I.~Manin,
``Construction Of Instantons,''
Phys.\ Lett.\ A {\bf 65} (1978) 185.
%%CITATION = PHLTA,A65,185;%%


\bibitem{Sem}
A.~M.~Semikhatov,
``Supersymmetric Instanton,''
JETP Lett.\  {\bf 35} (1982) 560
[Pisma Zh.\ Eksp.\ Teor.\ Fiz.\  {\bf 35} (1982) 452]; 
%%CITATION = JTPLA,35,560;%%
\\
A.~M.~Semikhatov,
``The Supersymmetric Instanton,''
Phys.\ Lett.\ B {\bf 120} (1983) 171.
%No rec. in SPIRES

\bibitem{Vo}
I.~V.~Volovich,
``Superselfduality For Supersymmetric Yang-Mills Theory,''
Phys.\ Lett.\ B {\bf 123} (1983) 329;
%%CITATION = PHLTA,B123,329;%%
\\ 
I.~V.~Volovich,
``On Superselfduality Equations,''
Theor.\ Math.\ Phys.\  {\bf 54} (1983) 55
[Teor.\ Mat.\ Fiz.\  {\bf 54} (1983) 89].
%%CITATION = TMPHA,54,55;%%


\bibitem{McAr1}
I.~N.~McArthur,
``Green Functions In A Superselfdual Yang-Mills Background,''
Nucl.\ Phys.\ B {\bf 239} (1984) 93.
%%CITATION = NUPHA,B239,93;%%

\bibitem{McAr2}
I.~N.~McArthur,
``Left Chiral 'Zero Modes' In A Superselfdual Yang-Mills Background,''
Nucl.\ Phys.\ B {\bf 241} (1984) 99.
%%CITATION = NUPHA,B241,99;%%

\bibitem{Si}
W.~Siegel,
``Supermulti - instantons in conformal chiral superspace,''
Phys.\ Rev.\ D {\bf 52} (1995) 1042
[arXiv:hep-th/9412011].
%%CITATION = HEP-TH 9412011;%%

\bibitem{GiMaReTa}
C.~R.~Gilson, I.~Martin, A.~Restuccia and J.~G.~Taylor,
``Selfduality In Superyang-Mills Theories,''
Commun.\ Math.\ Phys.\  {\bf 107} (1986) 377.
%%CITATION = CMPHA,107,377;%%

\bibitem{DeOg1}
C.~Devchand and V.~Ogievetsky,
``Super self duality as analyticity in harmonic superspace,''
Phys.\ Lett.\ B {\bf 297} (1992) 93
[arXiv:hep-th/9209120].
%%CITATION = HEP-TH 9209120;%%

\bibitem{DeOg2}
C.~Devchand and V.~Ogievetsky,
``The Matreoshka of supersymmetric selfdual theories,''
Nucl.\ Phys.\ B {\bf 414} (1994) 763
[arXiv:hep-th/9306163].
%%CITATION = HEP-TH 9306163;%%


\bibitem{NAC}
H.~Ooguri and C.~Vafa,
``The C-deformation of gluino and non-planar diagrams,''
Adv.\ Theor.\ Math.\ Phys.\  {\bf 7} (2003) 53
[arXiv:hep-th/0302109]; 
%%CITATION = HEP-TH 0302109;%%
\\ 
%H.~Ooguri and C.~Vafa,
%``Gravity induced C-deformation,''
%Adv.\ Theor.\ Math.\ Phys.\  {\bf 7} (2004) 405
%[arXiv:hep-th/0303063]; 
%%%CITATION = HEP-TH 0303063;%%
%\\
J.~de Boer, P.~A.~Grassi and P.~van Nieuwenhuizen,
``Non-commutative superspace from string theory,''
arXiv:hep-th/0302078;
%%CITATION = HEP-TH 0302078;%%
\\
N.~Seiberg,
``Noncommutative superspace, ${\cal N} = 1/2$ supersymmetry, 
field theory and  string theory,''
JHEP {\bf 0306} (2003) 010
[arXiv:hep-th/0305248]; 
%%CITATION = HEP-TH 0305248;%%
\\ 
N.~Berkovits and N.~Seiberg,
``Superstrings in graviphoton background 
and ${\cal N} = 1/2 + 3/2$ supersymmetry,''
JHEP {\bf 0307} (2003) 010
[arXiv:hep-th/0306226].
%%CITATION = HEP-TH 0306226;%%

\bibitem{LuZa}
J.~Lukierski and W.~J.~Zakrzewski,
``Euclidean Supersymmetrization Of Instantons And Selfdual Monopoles,''
Phys.\ Lett.\ B {\bf 189} (1987) 99.
%%CITATION = PHLTA,B189,99;%%

\bibitem{NiWa}
P.~van Nieuwenhuizen and A.~Waldron,
``On Euclidean spinors and Wick rotations,''
Phys.\ Lett.\ B {\bf 389} (1996) 29
[arXiv:hep-th/9608174].
%%CITATION = HEP-TH 9608174;%%

\bibitem{BeVaNi}
A.~V.~Belitsky, S.~Vandoren and P.~van Nieuwenhuizen,
``Instantons, Euclidean supersymmetry and Wick rotations,''
Phys.\ Lett.\ B {\bf 477} (2000) 335
[arXiv:hep-th/0001010].
%%CITATION = HEP-TH 0001010;%%

\bibitem{IvLeZu}
E.~Ivanov, O.~Lechtenfeld and B.~Zupnik,
``Nilpotent deformations of ${\cal N} = 2$ superspace,''
JHEP {\bf 0402} (2004) 012
[arXiv:hep-th/0308012].
%%CITATION = HEP-TH 0308012;%%

\bibitem{WeBa}
J.~Wess and J.~Bagger,
``Supersymmetry and supergravity,''
Princeton University Press, 1992. 

\bibitem{FermionZeroMode}
E.~Corrigan, D.~B.~Fairlie, S.~Templeton and P.~Goddard,
``A Green's Function For The General Selfdual Gauge Field,''
Nucl.\ Phys.\ B {\bf 140} (1978) 31; \\ 
%%CITATION = NUPHA,B140,31;%%
E.~Corrigan, P.~Goddard and S.~Templeton,
``Instanton Green Functions And Tensor Products,''
Nucl.\ Phys.\ B {\bf 151} (1979) 93.
%%CITATION = NUPHA,B151,93;%%


\bibitem{GrSoWe}
R.~Grimm, M.~Sohnius and J.~Wess,
``Extended Supersymmetry And Gauge Theories,''
Nucl.\ Phys.\ B {\bf 133} (1978) 275.
%%CITATION = NUPHA,B133,275;%%

\bibitem{So}
M.~F.~Sohnius,
``Bianchi Identities For Supersymmetric Gauge Theories,''
Nucl.\ Phys.\ B {\bf 136} (1978) 461.
%%CITATION = NUPHA,B136,461;%%


\bibitem{Fe} 
A.~Ferber,
``Supertwistors And Conformal Supersymmetry,''
Nucl.\ Phys.\ B {\bf 132} (1978) 55.

\bibitem{Ma}
Y.~I.~Manin,
``Gauge Field Theory And Complex Geometry,''
Springer, 1988. 

\bibitem{DInst}
A.~Imaanpur,
``On instantons and zero modes of ${\cal N} = 1/2$ SYM theory,''
 {\bf 0309} (2003) 077
[arXiv:hep-th/0308171]; 
%%CITATION = HEP-TH 0308171;%%
%A.~Imaanpur,
``Comments on gluino condensates in ${\cal N} = 1/2$ SYM theory,''
 {\bf 0312} (2003) 009
[arXiv:hep-th/0311137]; \\ 
%%CITATION = HEP-TH 0311137;%%
P.~A.~Grassi, R.~Ricci and D.~Robles-Llana,
``Instanton calculations for ${\cal N} = 1/2$ super Yang-Mills theory,''
 {\bf 0407} (2004) 065
[arXiv:hep-th/0311155]; \\ 
%%CITATION = HEP-TH 0311155;%%
R.~Britto, B.~Feng, O.~Lunin and S.~J.~Rey,
``${\rm U}(N)$ instantons on ${\cal N} = 1/2$ superspace: 
Exact solution and geometry of  moduli space,''
Phys.\ Rev.\ D {\bf 69} (2004) 126004
[arXiv:hep-th/0311275]; \\ 
%%CITATION = HEP-TH 0311275;%%
M.~Billo, M.~Frau, I.~Pesando and A.~Lerda,
``${\cal N} = 1/2$ gauge theory and its instanton moduli space 
from open strings in R-R background,''
 {\bf 0405} (2004) 023
[arXiv:hep-th/0402160]; \\ 
%%CITATION = HEP-TH 0402160;%%
S.~Giombi, R.~Ricci, D.~Robles-Llana and D.~Trancanelli,
``Instantons and matter in ${\cal N} = 1/2$ supersymmetric gauge theory,''
arXiv:hep-th/0505077.
%%CITATION = HEP-TH 0505077;%%

\bibitem{ATW2}
T.~Araki, T.~Takashima and S.~Watamura, 
work in progress.  


\end{thebibliography}
\end{document}